\documentclass{iau}  

\usepackage{graphicx,natbib,url,twoopt}
\usepackage{hyperref}                
\usepackage{pdfcomment}              
\usepackage{acronym}                 
\hypersetup{
  colorlinks=true,   
  urlcolor=blue,     
  linkcolor=red,     
}

\bibpunct{(}{)}{;}{a}{}{,}    

\def\RR #1\par{\noindent{\small \color{red} $\sharp$RR #1}\\[1ex]}
 
\bibpunct{(}{)}{;}{a}{}{,}    
\makeatletter
 \newcommandtwoopt{\citeads}[3][][]{%
   \nonstopmode
   \href{http://adsabs.harvard.edu/abs/#3}%
        {\def\hyper@linkstart##1##2{}%
         \let\hyper@linkend\@empty\citealp[#1][#2]{#3}}
   \biblink{#3}{\href{http://adsabs.harvard.edu/abs/#3}{ADS}}%
   \errorstopmode}            
 \newcommandtwoopt{\citepads}[3][][]{%
   \nonstopmode
   \href{http://adsabs.harvard.edu/abs/#3}%
        {\def\hyper@linkstart##1##2{}%
         \let\hyper@linkend\@empty\citep[#1][#2]{#3}}
   \biblink{#3}{\href{http://adsabs.harvard.edu/abs/#3}{ADS}}%
   \errorstopmode}            
 \newcommandtwoopt{\citetads}[3][][]{%
   \nonstopmode
   \href{http://adsabs.harvard.edu/abs/#3}%
        {\def\hyper@linkstart##1##2{}%
         \let\hyper@linkend\@empty\citet[#1][#2]{#3}}
   \biblink{#3}{\href{http://adsabs.harvard.edu/abs/#3}{ADS}}%
   \errorstopmode}            
 \newcommandtwoopt{\citeyearads}[3][][]{%
   \nonstopmode
   \href{http://adsabs.harvard.edu/abs/#3}%
        {\def\hyper@linkstart##1##2{}%
         \let\hyper@linkend\@empty\citeyear[#1][#2]{#3}}
   \biblink{#3}{\href{http://adsabs.harvard.edu/abs/#3}{ADS}}%
   \errorstopmode}            
\makeatother

\def\linkadspage#1#2#3{\href{http://adsabs.harvard.edu/cgi-bin/nph-data_query?bibcode=#1\&link_type=ARTICLE\&d_key=AST\#page=#2}{#3}}

\def\linkarxivpage#1#2#3{\href{http://lanl.arxiv.org/pdf/#1\#page=#2}{#3}}

\newcount\longrefs
\def\aap{\ifnum\longrefs=1 {Astron.\ Astrophys.}\else 
                           {A\hbox{\rm \&}A}\fi}
\def\aapr{\ifnum\longrefs=1 {Astron.\ Astrophys.\ Rev.}\else 
                            {A\hbox{\rm \&}AR}\fi}
\def\aaps{\ifnum\longrefs=1 {Astron.\ Astrophys.\ Suppl.}\else 
                            {A\hbox{\rm \&}A Suppl.}\fi}
\def\actaa{\ifnum\longrefs=1 {Acta Astronomica}\else
                            {Acta Astron.}\fi}
\def\aipcs{\ifnum\longrefs=1 {Am.\ Inst.\ Phys.\ Conf.\ Series}\else
                             {AIP Conf.\ Ser.}\fi}
\def\aj{\ifnum\longrefs=1 {Astron.\ J.}\else 
                          {AJ}\fi} 
\def\ao{\ifnum\longrefs=1 {Applied Optics}\else 
                           {Appl.\ Opt.}\fi} 
\def\aspcs{\ifnum\longrefs=1 {Astron.\ Soc.\ Pacific Conf.\ Series}\else 
                           {ASP Conf.\ Ser.}\fi} 
\def\apj{\ifnum\longrefs=1 {Astrophys.\ J.}\else 
                           {ApJ}\fi} 
\def\apjl{\ifnum\longrefs=1 {Astrophys.\ J. Lett.}\else 
                            {ApJL}\fi} 
\def\aplett{\ifnum\longrefs=1 {Astrophys.\ J. Lett.}\else 
                            {ApJ}\fi} 
\def\apjs{\ifnum\longrefs=1 {Astrophys.\ J. Suppl.}\else 
                            {ApJS}\fi}
\def\apss{\ifnum\longrefs=1 {Astrophys.\ and Space Science}\else 
                            {Astrophys.\ Space Sci.}\fi}
\def\araa{\ifnum\longrefs=1 {Ann.\ Rev.\ Astron.\ Astrophys.}\else 
                            {ARA\hbox{\rm \&}A}\fi}
\def\azh{\ifnum\longrefs=1 {Astronomicheskii Zhurnal}\else 
                            {Astron.\ Zhur.}\fi}
\def\baas{\ifnum\longrefs=1 {Bull.\ Am.\ Astron.\ Soc.}\else 
                            {BAAS}\fi}
\def\bain{\ifnum\longrefs=1 {Bull.\ Astronom.\ Institutes Netherlands}\else
                            {Bull.\ Astr.\ Inst.\ Neth.}\fi}
\def\cjaa{\ifnum\longrefs=1 {Chinese Jour.\ Astron.\ Astrophys.}\else 
                            {Chin.\ J.\ A\&A}\fi}
\def\gca{\ifnum\longrefs=1 {Geochim.\ Cosmochim.\ Acta}\else 
                           {Geochim.\ Cosmochim.\ Acta}\fi}
\def\grl{\ifnum\longrefs=1 {Geophys.\ Res.\ Lett.}\else 
                           {Geoph.\ Res.\ Lett.}\fi}
\def\iaucirc{\ifnum\longrefs=1 {IAU Circulars}\else 
                          {IAU Circ.}\fi}
\def\icarus{\ifnum\longrefs=1 {Icarus}\else 
                          {Icarus}\fi}
\def\ip{\ifnum\longrefs=1 {in press}\else 
                          {in press}\fi}
\def\jcap{\ifnum\longrefs=1 {Jour.\ Cosmology Astropart.\ Phys.}\else 
                          {JCAP}\fi}
\def\jgr{\ifnum\longrefs=1 {J.\ Geophys.\ Res.}\else 
                           {J.\ Geophys.\ Res.}\fi}  
\def\jrasc{\ifnum\longrefs=1 {J.\ Royal Astron.\ Soc.\ Canada}\else 
                           {JRAS Can.}\fi}  
\def\memsai{\ifnum\longrefs=1 {Mem.~Soc.~Astron.~Italiana}\else
                              {MmSAI}\fi}
\def\mnras{\ifnum\longrefs=1 {Mon.\ Not.\ Roy.\ Astron.\ Soc.}\else 
                             {MNRAS}\fi} 
\def\na{\ifnum\longrefs=1 {New Astronomy}\else 
                           {New Astron.}\fi}
\def\nar{\ifnum\longrefs=1 {New Astronomy rev.}\else 
                           {New Astron.\ Rev.}\fi}
\def\nat{\ifnum\longrefs=1 {Nature}\else 
                           {Nat}\fi}
\def\pasa{\ifnum\longrefs=1 {Pub.\ Astron.\ Soc.\ Australia}\else 
                            {PASA}\fi} 
\def\pasj{\ifnum\longrefs=1 {Pub.\ Astron.\ Soc.\ Japan}\else 
                            {PASJ}\fi} 
\def\pasp{\ifnum\longrefs=1 {Pub.\ Astron.\ Soc.\ Pacific}\else 
                            {PASP}\fi} 
\def\physscr{\ifnum\longrefs=1 {Physica Scripta}\else 
                            {Phys.\ Scrip.}\fi} 
\def\planss{\ifnum\longrefs=1 {Planetary \& Space Science}\else 
                            {Plan. \& Space Sci.}\fi} 
\def\procspie{\ifnum\longrefs=1 {Proc.\ SPIE}\else 
                            {Proc.\ SPIE}\fi} 
\def\qjras{\ifnum\longrefs=1 {Quarterly J.\ Royal Astron.\ Soc.}\else 
                            {QJRAS}\fi} 
\def\rmxaa{\ifnum\longrefs=1 {Revista Mexicana de Astron.\ y Astrofys.}\else 
                            {RMxAA}\fi} 
\def\sa{\ifnum\longrefs=1 {Soviet Astron..}\else 
                               {Sov.\ Astron.}\fi}
\def\skytel{\ifnum\longrefs=1 {Sky \& Telescope}\else 
                            {Sky \& Tel.}\fi} 
\def\solphys{\ifnum\longrefs=1 {Solar Phys.}\else 
                               {SoPh}\fi}
\def\sovast{\ifnum\longrefs=1 {Soviet Astronomy}\else 
                               {Sov.\ Ast.}\fi}
\def\ssr{\ifnum\longrefs=1 {Space Science Rev.}\else 
                               {Space\ Sci.\ Rev.}\fi}
\def\zap{\ifnum\longrefs=1 {Zeitschr.\ f.\ Astrophysik}\else
                               {Z.\ Astrophys.}\fi}

\makeatletter
\newcommand{\bibnote}[2]{\global\@namedef{#1note}{#2}}
\newcommand{\biblink}[2]{\global\@namedef{#1link}{#2}}
\makeatother

\def\wlRRssx#1{\href{http://www.staff.science.uu.nl/~rutte101/rrweb/rjr-edu/lectures/rutten_ssx_lec.pdf}{#1}}

\def\RTSA{\href{http://adsabs.harvard.edu/abs/2003rtsa.book.....R}%
{RTSA}} 
\def\RTSAp#1#2{\href{http://www.staff.science.uu.nl/~rutte101/rrweb/rjr-edu/coursenotes/rutten_rtsa_notes_2003.pdf\#page=#1}{#2}}

\newacro{AA}{Astronomy \& Astrophysics}  
\newacro{ADS}{Astrophysics Data System}
\newacro{AIA}{Atmospheric Imaging Assembly}
\newacro{ALMA}{Atacama Large Millimeter/submillimeter Array}
\newacro{AO}{adaptive optics}
\newacro{ApJ}{Astrophysical Journal}
\newacro{AR}{active region}
\newacro{bb}{bound-bound}
\newacro{bf}{bound-free}
\newacro{BFI}{Broad-band Filter Imager}
\newacro{CE}{coronal equilibrium}
\newacro{CfA}{Center for Astrophysics}
\newacro{CME}{coronal mass ejection}
\newacro{CRD}{complete redistribution}
\newacro{CRISP}{CRisp Imaging SpectroPolarimeter}
\newacro{CRISPEX}{CRisp SPectral EXplorer}
\newacro{CS}{coherent scattering}
\newacro{DEM}{Differential Emission Measure}
\newacro{DF}{dynamic fibril}
\newacro{DKIST}{Daniel K. Inouye Solar Telescope}
\newacro{DLR}{Deutsches Zentrum f\"ur Luft- und Raumfahrt}
\newacro{DOT}{Dutch Open Telescope}
\newacro{DST}{Richard B. Dunn Solar Telescope}   
\newacro{EB}{Ellerman bomb}
\newacro{EDP}{\'{E}dition Diffusion Presse}  
\newacro{EIT}{Extreme ultraviolet Imaging Telescope}
\newacro{EPIC}{European participation in Solar-C}
\newacro{ERC}{European Research Council}
\newacro{ESA}{European Space Agency}
\newacro{EST}{European Solar Telescope}
\newacro{EUV}{extreme ultraviolet}
\newacro{FAF}{flaring active-region fibril}
\newacro{ff}{free-free}
\newacro{FITS}{Flexible Image Transport System}
\newacro{FOV}{field of view}
\newacro{fov}{field of view}
\newacro{FWHM}{full width at half maximum}
\newacro{HAO}{High Altitude Observatory}
\newacro{HD}{hydrodynamics}
\newacro{Hi-C}{High Resolution Coronal Imager Sounding Rocket}
\newacro{HMI}{Helioseismic and Magnetic Imager}
\newacro{IAA}{Instituto de Astrof\'{i}sica de Andaluc\'{i}a}
\newacro{IAC}{Instituto de Astrof\'{i}sica de Canarias}
\newacro{IAS}{Institut d'Astrophysique Spatiale}
\newacro{IAU}{International Astronomical Union}
\newacro{IBIS}{Interferometric Bi-dimensional Spectrometer}
\newacro{IDL}{Interactive Data Language}
\newacro{IMaX}{Imaging Magnetograph eXperiment}
\newacro{INAF}{Istituto Nazionale di Astrofisica}
\newacro{IB}{IRIS bomb}
\newacro{IR}{infrared}
\newacro{IRIS}{Interface Region Imaging Spectrograph}
\newacro{ISAS}{Institute of Space and Astronautical Science}
\newacro{ISP}{Institute for Solar Physics}
\newacro{ISS}{International Space Station}
\newacro{ISSI}{International Space Science Institute}
\newacro{ITA}{Institute for Theoretical Astrophysics}
\newacro{JAXA}{Japan Aerospace Exploration Agency}
\newacro{KIS}{Kiepenheuer--Institut f\"{u}r Sonnenphysik}
\newacro{KPNO}{Kitt Peak National Observatory}
\newacro{LASP}{Laboratory for Atmospheric and Space Physics}
\newacro{LC}{liquid cristal}
\newacro{LMSAL}{Lockheed Martin Solar and Astrophysics Labratory}
\newacro{LOS}{line of sight}
\newacro{LTE}{local thermodynamic equilibrium}
\newacro{MC}{magnetic concentration}
\newacro{MCAO}{multi-conjugate adaptive optics} 
\newacro{MDI}{Michelson Doppler Imager}
\newacro{ME}{Milne-Eddington} 
\newacro{MHD}{magnetohydrodynamics}
\newacro{MOMFBD}{Multi-Object Multi-Frame Blind Deconvolution}
\newacro{MPE}{Max--Planck--Institut f\"ur extraterrestrische Physik}
\newacro{MPG}{Max--Planck--Gesellschaft}
\newacro{MPS}{Max Planck Institute for Solar System Research}
\newacro{MSSL}{Mullard Space Science Laboratory}
\newacro{MTF}{modulation transfer function}
\newacro{NAOJ}{National Astronomical Observatory of Japan}
\newacro{NASA}{National Aeronautics and Space Administration}
\newacro{NIST}{National Institute of Standards and Technology}
\newacro{NLTE}{non-local thermodynamic equilibrium}
\newacro{NLFFF}{non-linear force-free field}
\newacro{NOAA}{National Oceanic and Atmospheric Administration}
\newacro{non-E}{non-equilibrium}
\newacro{NSO}{National Solar Observatory}
\newacro{NWO}{Netherlands Organisation for Scientific Research}
\newacro{PHE}{propagating heating event}
\newacro{PRD}{partial redistribution}
\newacro{PROBA2}{PRoject for OnBoard Autonomy}
\newacro{PSBE}{post Saha-Boltzmann extinction}
\newacro{PSF}{point spread function}
\newacro{QS}{quiet Sun}
\newacro{QSEB}{quiet-Sun Ellerman-like brightening} 
\newacro{RAL}{Rutherford Appleton Laboratory}
\newacro{RBE}{rapid blue-shifted excursion}
\newacro{R-MHD}{radiation hydrodynamics}
\newacro{rms}{root mean square}
\newacro{RMS}{root mean square}
\newacro{ROB}{Royal Observatory of Belgium}
\newacro{ROI}{region of interest}
\newacro{RRE}{rapid red-shifted excursion}
\newacro{RTE}{radiative transfer equation}
\newacro{RTSA}{Radiative Transfer in Stellar Atmospheres}
\newacro{SE}{statistical equilibrium}
\newacro{SB}{Saha Boltzmann}
\newacro{SDO}{Solar Dynamics Observatory}
\newacro{SJI}{slit-jaw image}
\newacro{SNR}{signal-to-noise ratio}
\newacro{SO}{Solar Orbiter}
\newacro{SoHO}{Solar and Heliospheric Observatory}
\newacro{SP}{Spectropolarimeter}
\newacro{SST}{Swedish 1-m Solar Telescope}
\newacro{SUMER}{Solar Ultraviolet Measurements of Emitted Radiation}
\newacro{SUFI}{Sunrise Filter Imager}
\newacro{SVD}{singular value decomposition}
\newacro{SVST}{Swedish Vacuum Solar Telescope}
\newacro{THEMIS}{T\'{e}lescope H\'{e}liographique pour l'Etude du 
   Magn\'{e}tisme et des Instabilit\'{e} Solaires}     
\newacro{TR}{transition region}
\newacro{TRACE}{Transition Region and Coronal Explorer}
\newacro{TSI}{total solar irradiance}
\newacro{UT}{Universal Time}
\newacro{UV}{ultraviolet}
\newacro{VAULT}{Very high angular resolution ultraviolet telescope}
\newacro{VIRGO}{Variability of solar IRradiance and Gravity Oscillations}
\newacro{VTT}{Vacuum Tower Telescope}    
\newacro{XRT}{X-Ray Telescope}

\def\acp#1{\pdftooltip{\acs{#1}}{\acl{#1}}} 

\def\nl{,\ } 

\def\LA{Lingezicht Astrophysics\nl 't Oosteneind 9\nl 4158\,CA Deil\nl 
        The Netherlands}

\def\NAOJ{National Astronomical Observatory of Japan\nl
          2-21-1 Osawa, Mitaka\nl Tokyo 181-8588\nl Japan}

\long\def\startignore #1\stopignore{}   

\def\rmit#1{{\it #1}}              
           
\def\ie{\rmit{i.e.,}}              
\def\eg{\rmit{e.g.,}}              
\def\cf{cf.}                       

\def\specchar#1{\uppercase{#1}}    
\def\specand{ and }                
\def\specand{\,\&\,}               
\def\AlI{\mbox{Al\,\specchar{i}}}  

\def\CII{\mbox{C\,\specchar{ii}}}

\def\FeI{\mbox{Fe\,\specchar{i}}}

\def\HI{\mbox{H\,\specchar{i}}} 
 
\def\Hmin{\hbox{{\rm H}$^{^{_-}}$}}      

\def\MgI{\mbox{Mg\,\specchar{i}}}

\def\NaI{\mbox{Na\,\specchar{i}}}

\def\SiI{\mbox{Si\,\specchar{i}}}

\def\SiIV{\mbox{Si\,\specchar{iv}}}

\def\Halpha{\mbox{H\hspace{0.1ex}$\alpha$}} 

\def\Lyalpha{\mbox{Ly$\hspace{0.2ex}\alpha$}}
\def\Lybeta{\mbox{Ly$\hspace{0.2ex}\beta$}}

\def\NaID{\mbox{Na\,\specchar{i}\,\,D}}
\def\NaIDone{\mbox{Na\,\specchar{i}\,\,D$_1$}}

\def\CaIIK{\mbox{Ca\,\specchar{ii}\,\,K}}       

\def\CaIIHK{\mbox{Ca\,\specchar{ii}\,\,H{\specand}K}}

\def\CaIR{\mbox{Ca\,\specchar{ii}\,8542\,\AA}} 

\def\MgIIk{\mbox{Mg\,\specchar{ii}\,\,k}}

\def\level #1 #2#3#4{$#1 \; ^{#2} \mbox{#3} ^{#4}$}   

\def\rmb{{\rm b}}     
  
 \def\rmD{{\rm D}} 
\def\rme{{\rm e}}

 \def\rmH{{\rm H}}

 \def\rmK{{\rm K}}

\def\tis{\!=\!}                            
\def\tapprox{\!\approx\!}                  

\def\linkcutssx#1#2{\href{http://www.staff.science.uu.nl/~rutte101/rrweb/rjr-pubstuff/lyalma/cutssx.pdf\#page=#1}{#2}}  
\def\linkcutssxp#1#2{\href{http://www.staff.science.uu.nl/~rutte101/rrweb/rjr-pubstuff/lyalma/cutssx.pdf\#page=#1}{display #1 = #2}}  

\def\linkssf#1{\href{https://www.staff.science.uu.nl/~rutte101/rrweb/rjr-edu/lectures/rutten_ssf_lec.pdf}{#1}}  

\def\parrr#1{\par \vspace{1.2ex} \noindent {\em #1.}~}  

\title{Solar ALMA predictions: tutorial}
\author[Robert J. Rutten]{Robert J. Rutten}
\affiliation{\LA\\ 
 \NAOJ\\ email: {\tt R.J.Rutten@uu.nl}}
\pubyear{2017} \volume{327} 
\jname{Fine structure and dynamics of the solar atmosphere} 
\editors{Eds: Santiago Vargas Dom\'{\i}nguez, Alexander Kosovichev,\\
   Louise Harra, Patrick Antolin}

\begin{document}  

\maketitle

\begin{abstract}
I have proposed that long \Halpha\ fibrils are caused by 
heating events of which the tracks are afterwards outlined by
contrails of cooling gas with extraordinary \Halpha\ opacity and yet
larger opacity at the \acp{ALMA} wavelengths.
Here I detail the radiative transfer background.
\end{abstract}

\firstsection
\section{Introduction}     \label{sec:introduction}
\noindent 
Recently I proposed that many long \Halpha\ fibrils, \ie\ the slender
\Halpha\ features that extend from plage and network far out over
adjacent quiet internetwork areas wherever the Sun shows some
activity, are contrails 
(\citeads{2016arXiv160901122R},  
\def\PubI{\linkarxivpage{1609.01122}{1}{Pub~1}}  
henceforth \PubI).
Airplane contrails on our sky are water recondensation features
tracking sudden heating in jet engines. 
\Halpha\ contrails on the solar disk are hydrogen recombination
features tracking sudden heating in jet-like heating events.
A striking example with precursor heating to very high temperature was
detailed by \citetads{2016arXiv160907616R}. 

Observationally, I based this proposal not only on the latter example
but also on observed paucity of internetwork shock scenes in \Halpha,
large disalikeness between \Halpha\ and \Lyalpha\ images, and
fibril incongruity between \Halpha\ and \CaIR. 

Theoretically, I based this proposal on appreciating how \Lyalpha\
controls hydrogen populations in neutral-hydrogen gas and so produces
non-equilibrium (non-E) ionization-recombination balancing in cooling
hydrogen gas.

A corollary is that such \Halpha\ contrails are as opaque or much more
opaque at the \acp{ALMA} wavelengths. 
\linkarxivpage{1609.01122}{9}{Section~5} of \PubI\ gives a dozen
explicit predictions of what \acp{ALMA} may therefore see and not see
on the Sun. 
In a nutshell, I predict that internetwork areas are covered by opaque
canopies wherever \Halpha\ shows fibrils, but as umbrellas rather than
fibrilar canopies, \ie\ opaque blankets with less lateral contrast.
These will obscure not only activity features underneath such as
Ellerman bursts (\citeads{1917ApJ....46..298E}; 
review by \citeads{2013JPhCS.440a2007R}), 
but also quiet-Sun features as the predicted shock scenes in \eg\
\linkadspage{2016Msngr.163...15W}{4}{Fig.~4} of
\citetads{2016Msngr.163...15W} 
-- unless we sink in a Maunder minimum.  

I presented the above at the symposium.
Here I detail radiative transfer background that was only summarized
in \linkarxivpage{1609.01122}{2}{Sect.~2.1} of \PubI\ by linking to a
\linkcutssx{1}{selection} of my \wlRRssx{solar spectrum example
displays} as a shortcut to avoid inclusion of more figures.
I show these figures here and explain them with references to my
\acp{ADS}-available course notes
(\citeads{2003rtsa.book.....R}, 
henceforth \RTSAp{1}{RTSA}),
which are largely a didactic rendering of material covered already by
\citetads{1970stat.book.....M}. 
Although optically thick radiative transfer is long-standing textbook
material, it remains intransparent. 
I therefore use graphical exposition here in the order \acp{LTE} --
\acp{NLTE} -- \acp{non-E}. 
I also treat Rydberg lines (skipped in \PubI).\footnote{Links to cited
figures and equations open the pertinent page in a browser depending
on your pdf viewer, its permission setup, and your publisher access. 
The name-year clickers open the \acp{ADS} abstract page.
Acronyms may show their meaning at mouse-over (not in Preview).}

\begin{figure*}[t]
  \centerline{\includegraphics[width=0.99\textwidth]{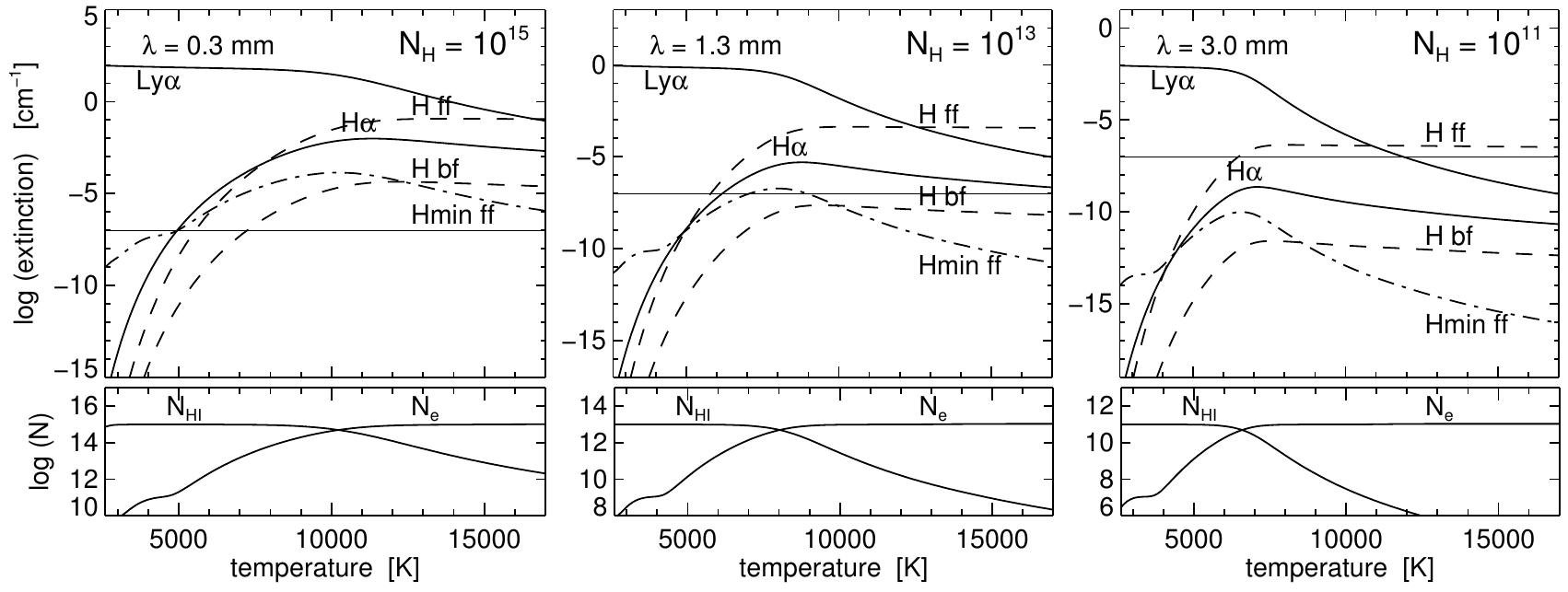}}
  \caption[]{\label{fig:SB} 
  Saha-Boltzmann line extinction coefficient against temperature at
  the centers of \Lyalpha\ and \Halpha\ (solid) and continuous
  extinction coefficient of the \HI\ free-free and bound-free
  contributions (dashed) and the \Hmin\ free-free contribution
  (dot-dashed) at three \acp{ALMA} wavelengths (from left to right
  0.35, 1.3 and 3.0\,mm), for gas of solar composition with different
  total hydrogen densities (from left to right
  $N_\rmH \tis 10^{15}, 10^{13}$ and $10^{11}$~cm$^{-3}$,
  corresponding to the radiation escape heights at these wavelengths
  at the bottom, middle and top of the ALC7 chromosphere).
  The horizontal line at $y \tis -7$ shows at which extinction a
  100\,km thick feature becomes optically thick.
  The small lower panels show the competing neutral hydrogen and
  electron densities (part\,cm$^{-3}$), at the same logarithmic unit
  (dex) size as the extinction scales to enable slope comparisons.   
  The $y$ scales shift between columns.
  Selection from \linkarxivpage{1609.01122}{4}{Fig.~1} of \PubI.
  }
\end{figure*}

\section{Hydrogen in Saha-Boltzmann equilibrium}    
\label{sec:SB} \noindent
\parrr{ALMA as thermometer}
\acp{ALMA} is often advertised as a linear thermometer of the solar
atmosphere, quoting ``\acp{LTE} formation of solar mm continua''.
\acp{LTE} holds strictly only for the source function of free-free
(ff) processes since these are collisional and the Maxwell
distribution is valid in chromospheric conditions.
However, $S \tapprox B$ is a good approximation for mm bound-free (bf)
continua which do not suffer the tremendous scattering with
$S \tapprox J$ that characterizes all strong solar lines and also all
\acp{bf} ultraviolet continua (examples below).  
The assumption is wrong for Thomson and Rayleigh scattering, but their
contributions are negligible in chromospheric conditions at mm
wavelengths, also unlike the optical and ultraviolet. 

Since solar mm radiation obeys the Rayleigh-Jeans approximation,
\acp{ALMA}-measured brightness temperature from optically thick
features therefore indeed represents electron temperature when
absolute intensity calibration is achieved.
The signal then represents $T(\tau_\mu \tis 1)$, \ie\ the temperature
at the representative Eddington-Barbier location defined by the summed
extinction along the line of sight. 
Optically thin features show (apart from their background)
$T_\rmb = \tau\,T$, a fraction of the temperature scaled by the summed
extinction along the beam.
The question therefore is one of extinction -- is it \acp{LTE},
\acp{NLTE}, \acp{non-E}, and how thick does it make a feature of
interest at mm wavelengths?
I start with \acp{LTE} and then treat \acp{NLTE} and \acp{non-E} which
require non-local multi-level scrutiny of hydrogen population
processes.

\parrr{Saha-Boltzmann opacities}
The definition of \acp{LTE} is local validity of Saha and Boltzmann
(SB) partitioning for the populations of all pertinent particle levels
and stages (\RTSAp{48}{Sect.~2.5} on page~28 of \RTSA). 
The extinction coefficient for any transition can then be evaluated
from the local temperature, electron density, and elemental abundance
without knowledge of impinging radiation fields, even if these do not
correspond to the local temperature.
Iterative solution is required because the electron density enters
reciprocally in the Saha population ratio between successive
ionization stages and depends on the degree of ionization of hydrogen,
helium, and abundant metals and on the degree of pertinent molecular
binding, in particular of H into H$_2$ (\RTSAp{163}{Sect.~7.2.2} on
page~143 of \RTSA; more detail in Sects.~3.1 and 3.2 of
\citeads{1970stat.book.....M}). 

The lower panels of Fig.~\ref{fig:SB} show resulting densities for gas
of solar composition. 
In the small plateau near 4000\,K the free electrons with
$N_\rme \approx 10^{-4} N_\rmH$ come from abundant low-ionization
metals (Mg, Fe, Si, Al); these govern photospheric \Hmin\ extinction.
At right virtually all free electrons come from hydrogen.

The \acp{SB} extinction curves in the upper panels of
Fig.~\ref{fig:SB} follow the hydrogen ionization. 
The decay slope for \Lyalpha\ is the same as for $N_{\rm H\!\,I}$,
but the decay of \Halpha\ is largely compensated by the increasing
Boltzmann ratio (for \acp{SB} \Halpha\ even exceeds \Lyalpha\ above
40\,000\,K from $g_2/g_1 \tis 8$).
The Boltzmann factor makes the initial rise for \Halpha\ exceedingly
steep, spanning over 10 dex (orders of magnitude).
\Halpha\ extinction is nearly negligible below 5000\,K, which explains
why the \CaIIHK\ lines show much larger dips in the photospheric
spectrum, but above 10\,000\,K the \acp{SB} extinction of \Halpha\
becomes much larger than that of \MgIIk\ and at high density even
comparable to that of the \CII\ and \SiIV\ resonance lines
(\linkadspage{2016A&A...590A.124R}{9}{Fig.~5} of
\citeads{2016A&A...590A.124R}). 

In Fig.~\ref{fig:SB} \Hmin\,\acp{ff} extinction is the main continuum
agent below 5000\,K while \HI\,\acp{ff} extinction, sharing the steep
Boltzmann increase, wins at higher temperature.
Therefore, across the parameter domain of Fig.~\ref{fig:SB} the
\acp{ff}-process $S \tis B$ equality indeed holds.

Comparison of these curves with the horizontal line at $y \tis -7$
shows that in \acp{SB} conditions a feature of 100\,km thickness and
temperature above 6000\,K is optically thick to extremely thick in
\Halpha\ and \HI\,\acp{ff} in the first column, still thick in the second,
thin in \Halpha\ but still thickish in \HI\,\acp{ff} in the third. 
The difference \HI\,\acp{ff}$-$\Halpha\ increases with wavelength because
$\alpha^{\rm ff}_\nu \sim \lambda^2 N_\rme N_{\rm ion}\,T^{-3/2}$
(\RTSAp{47}{Eq.~2.79} on page~27 of \RTSA).
Hence, if solar features have \acp{SB} opacities then their
visibility in \Halpha\ implies equal or better visibility at the
\acp{ALMA} wavelengths, more so at higher temperature and longer
wavelength.

\begin{figure*}[t]
  \centerline{\includegraphics[width=0.99\textwidth]{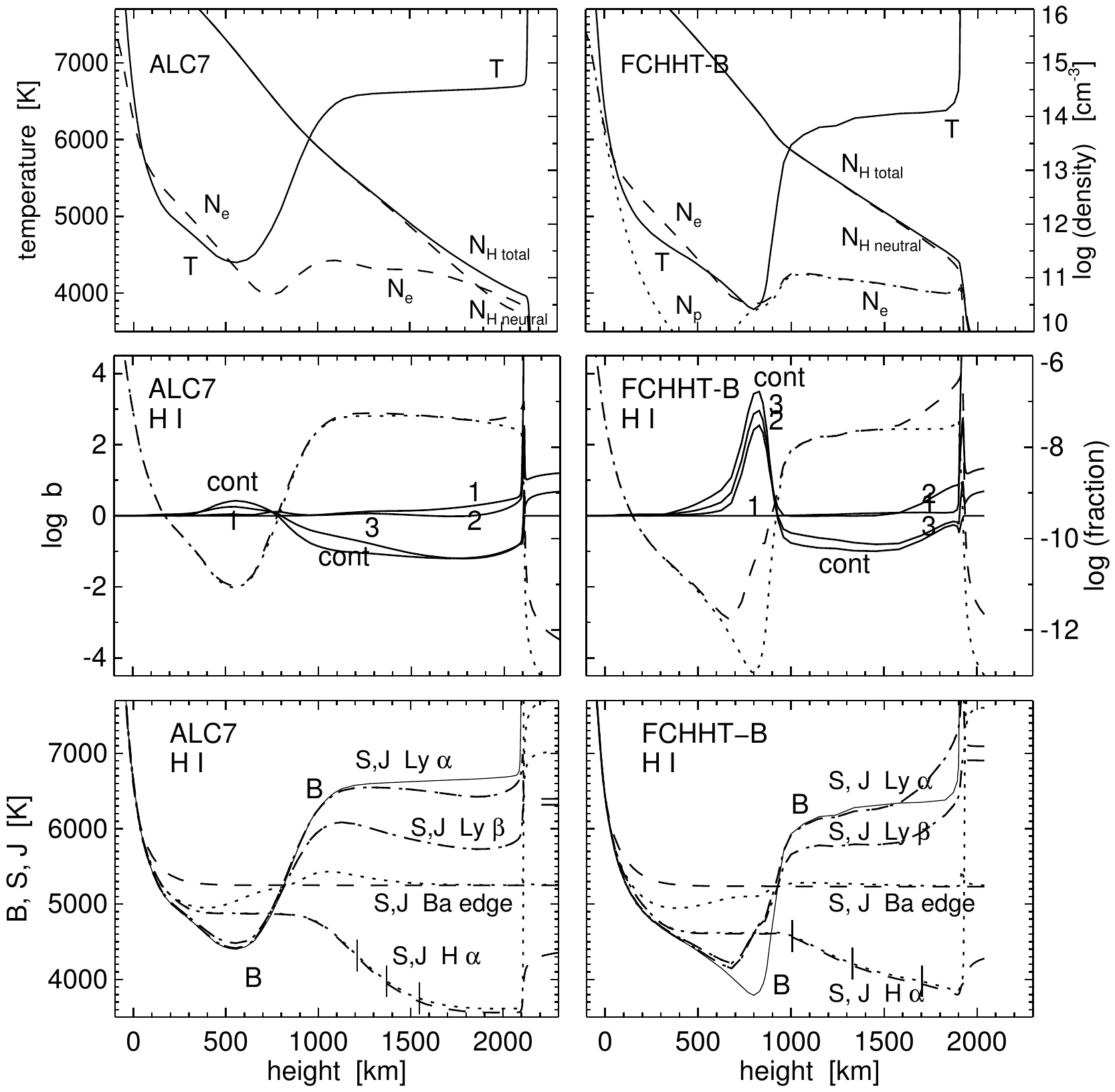}}
  \caption[]{\label{fig:hydmodels}
  Hydrogen in static plane-parallel atmospheres.
  Left: ALC7 of \citetads{2008ApJS..175..229A}. 
  Right: FCHHT-B of
  \citetads{2009ApJ...707..482F}. 
  Top row: model properties against height. 
  Density scale at right.
  Center row: \acp{NLTE} departure coefficients for hydrogen levels
  1--3 and the continuum.
  The dashed curves show the fractional population $n_2/N_\rmH$ of the
  $n \tis 2$ hydrogen level in \acp{NLTE}, the dotted curves in
  \acp{LTE}, with scale at right. 
  Bottom row: $B$, $S$ and $J$ for hydrogen features, plotted as
  representative temperatures.
  The ticks on the $S$ curves for \Halpha\ are for
  $\tau\tis 3, 1, 0.3$. 
  The Balmer edge curves are 100\,\AA\ averages shortward of 
  3646\,\AA\ including lines.
  }
\end{figure*}

\section{Hydrogen in static NLTE solar  model atmospheres}    
\label{sec:static}

\parrr{Standard models}
I now turn to hydrogen behavior in \acp{NLTE}, using the unrealistic
simplification of static plane-parallel modeling because hypothetical
idealized atmospheres of this type have the didactic virtue
of producing fully understandable spectra\footnote{I regard 1D models
as computationally existing didactically superb non-solar-like
plane-parallel stars with remarkably solar-like but misleading spectra
(\RTSAp{209}{page~189} of \RTSA).}. 
See \citetads{2002JAD.....8....8R} 
for a review of such modeling.

Figure~\ref{fig:hydmodels} shows two recent contenders.
I prefer ALC7 because its chromospheric extent is based on turbulent
pressure constrained by observed non-thermal line broadening,
similarly to the predecessor VALIII models of
\citetads{1981ApJS...45..635V} 
and FAL models of \citetads{1993ApJ...406..319F}, 
whereas the extent of FCHHT-B (without the A of Avrett) is based on
ad-hoc gravity modification in addition to this turbulent pressure.
The FCHHT-B model has a higher-located and appreciably cooler
temperature minimum, introduced to reproduce observed deep CO lines
that in reality are due to non-static \acp{non-E} cooling clouds
(\citeads{2000ApJ...536..481U}; 
\citeads{2003ApJ...588L..61A}). 
I add this model here because its steep temperature rise to its
chromosphere produces didactically valuable \Lyalpha\ back-radiation
discussed below. 
The near-isothermal chromospheres of both models are effectively
constrained by acoustic shocks in internetwork areas and represent
their temperatures, densities, and degree of hydrogen ionization
rather than quiet-Sun averages of these quantities
(\citeads{1995ApJ...440L..29C}). 

In both models the photospheric temperature gradient is close to
radiative equilibrium. 
The electron density follows $N_\rmH$ at the $10^{-4}$ offset governed
by low-ionization metals until hydrogen ionizes.
It then remains fairly constant throughout the model chromospheres,
and so is therefore the collisional destruction probability
$\varepsilon$ (\RTSAp{84}{Sect.~3.4} on page~64 of \RTSA) which scales
primarily with $N_\rme$ (\RTSAp{70}{Eq.~3.33} on page~50 of \RTSA). 
Since the temperature is also fairly constant these model
chromospheres are similar to the classic isothermal
constant-$\varepsilon$ atmosphere for which the theory of scattering
line formation was developed half a century ago, in particular in
\citetads{1965SAOSR.174..101A} 
(\RTSAp{112}{Sect.~4.3} on page~92\,ff of \RTSA).

\parrr{NLTE populations}
The center panels of Fig.~\ref{fig:hydmodels} show \acp{NLTE}
population departure coefficients\footnote{Beware: these are Zwaan
coefficients, not Menzel coefficients used by E.H.~Avrett and
J.~Fontenla. 
See \RTSAp{56}{warning} on page~36 of \RTSA\ regarding
misinterpretation, as indeed done by
\citetads{2009ApJ...707..482F} 
and diagnosed by
\citetads{2012A&A...540A..86R}.} 
$b \tis n/n^{\rm LTE}$ for the first three levels and the continuum
state of hydrogen (free protons).
Below the transition region the ground level has $b_1\tapprox 1$, as
expected since where hydrogen is mostly neutral almost all hydrogen
particles sit in that level.
The second level also shows near-\acp{SB} population, except in the
temperature minimum and the onset to the transition region. 
The third level has $b_3$ dropping towards the continuum offset
$b_{\rm cont} \tapprox 0.1$ across both chromospheres.
The FCHHT-B model has high $b$ peaks for the higher levels in its
temperature minimum.
Except for $b_1$ these behaviors have to do with the \Lyalpha,
\Halpha\ and Balmer-continuum source functions. 
These are the main topic here.

\parrr{Meaning of NLTE}
\acp{NLTE} means that \acp{LTE} cannot be assumed.
Usually it means that statistical equilibrium (SE) is assumed.
Often, it is taken to mean only what I call source function
\acp{NLTE}: $S \!\neq\! B$ -- but often for photospheric lines opacity
\acp{NLTE} is more important, \eg\ for \FeI\ lines
(\citeads{1988ASSL..138..185R} and below). 

\acp{NLTE} is primarily about scattering, \ie\ effects of
impinging radiation brought from elsewhere by scattering.
This non-local influence can indeed affect line source functions, in
the Wien approximation (valid for Lyman and Balmer transitions)
measured as $S^l \tapprox (b_u/b_l)\, B$, but also line extinction
coefficients, in the Wien approximation measured as
$\alpha^l \approx b_l \, \alpha^l_{\rm LTE}$ (\RTSAp{53}{Sect.~2.6.2}
on page~33 of \RTSA).
Source function \acp{NLTE} ($b_u/b_l \!\neq\! 1$) is usually due to
\acp{bb} scattering, opacity \acp{NLTE} ($b_l \!\neq\! 1$)
to \acp{bf} scattering.  
I showcase both.

\parrr{Bound-bound scattering}
The \acp{bb} line source function can be decomposed into distinct
contributions as
\begin{equation}
  S^l = (1-\varepsilon-\eta)\,J + \varepsilon\,B + \eta\,S^\rmD,
  \label{eq:Sl}
\end{equation}
weightedly summing different sources for producing new line photons:
scattering from the local photon reservoir (angle-averaged intensity
$J$), creation by collisional excitation using thermal energy described
by $B$, and detour production via lower-to-upper paths along other
levels and stages, with $S^\rmD$ a formal combined source function for
all such paths which may include ionization plus recombination
(\linkssf{teaching display ``all bb pairs''}).

This is a fundamental equation but ``not yet'' in \RTSA\ -- I believe
it has been published only by me so far in this form using
$\varepsilon$ (fraction of photoexcitations followed by direct
collisional upper-to-lower deexcitation) and $\eta$ (fraction of
photoexcitations followed by detour upper-to-lower paths). 
The classic literature used instead the ratios $\varepsilon^\prime$
and $\eta^\prime$ of such extinctions to the contribution by two-level
scattering following \citetads{1957ApJ...125..260T} 
(\eg\ \linkadspage{1968slf..book.....J}{199}{Eq.~8.5} on page~181 of
\citeads{1968slf..book.....J}). 

\begin{figure*}[t]
  \centerline{\includegraphics[width=0.99\textwidth]{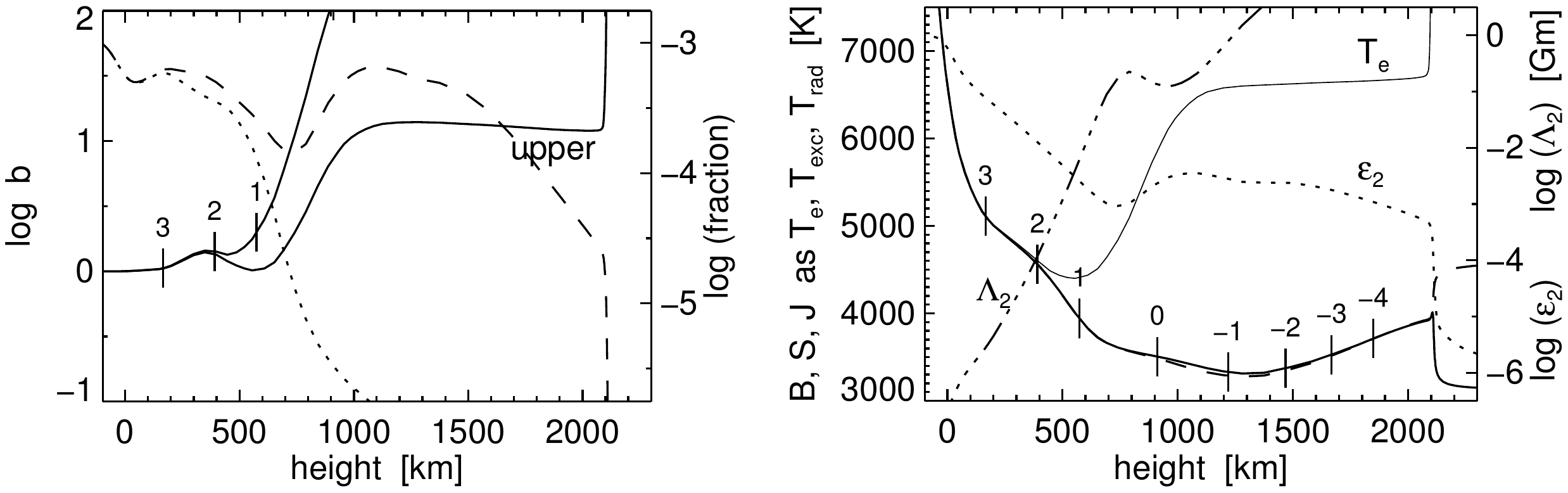}}
  \caption[]{\label{fig:NaD}
  Formation of \NaIDone\ in the ALC7 atmosphere. 
  Left: departure coefficients $b_l$ and $b_u$ (solid) and fractional
  lower-level population in \acp{NLTE} (dashed) and \acp{LTE} (dotted)
  with scale at right.  
  Right: $B$, $S$ (solid) and $J$ (dashed) represented as formal
  temperatures (electron, excitation, radiation defined on
  \RTSAp{57}{page~37\,ff} of \RTSA).
  Dotted: collisional destruction probability $\varepsilon$ in
  two-level approximation (ignoring detours). 
  Scale to the right.
  Dot-dashed: thermalization length for the Doppler core in Gigameter,
  same scale. 
  The label $\Lambda_2$ is placed near the line-core thermalization
  height.
  The numbered ticks show values of $\log \tau$ at line center;
  $\log \tau \tis 0$ represents the characteristic Eddington-Barbier
  sampling height for vertical viewing.
  }
\end{figure*}

Figure~\ref{fig:NaD} shows the formation of \NaIDone\ in ALC7 because
this line presents a clean example of \acp{bb} resonance scattering.
The detour terms are negligible for this line, and also the effects of
coherent scattering (partial redistribution, PRD) can be ignored
(\citeads{1992A&A...265..268U}). 
However, its $\varepsilon \tapprox 10^{-3}$ in the chromosphere means 
that resonance scattering dominates there.  
The line source function (at line center $\approx$ total source
function because the contribution by continuous extinction
processes is negligible,
\RTSAp{33}{Eq.~2.23} on page~13 of \RTSA)
follows the scattering decline illustrated by
\citetads{1965SAOSR.174..101A} 
to the small value at the escape surface given by the
``$\sqrt{\varepsilon}$\,'' law with
$I \approx S(\tau \tis 1) \approx S(\tau \tis 0) =
\sqrt{\varepsilon}\,S$ (\RTSAp{117}{Eq.~4.81} on page~97 of \RTSA).
Correspondingly, the thermalization depth is large:
$\Lambda \tapprox 1/\varepsilon$ in optical depth units for the
Doppler core and complete redistribution (\acp{CRD})
(\RTSAp{130}{Eq.~4.106} on page~110 of \RTSA). 
This is roughly the depth at which the radiation does not yet sense
the presence of a surface where it can leak out and therefore remains
controlled by the local temperature (Avrett's \RTSAp{129}{Fig.~4.13} on
page~109 of \RTSA). 

Their scattering makes the \NaID\ lines the darkest solar lines in the
optical spectrum. 
Figure~\ref{fig:NaD} shows that the \NaIDone\ core is chromospheric in
ALC7 in the sense that the last scattering encoding Dopplershift and
Zeeman signature takes place around 900\,km, but the core photons that
escape there were thermally created near 200\,km in the low ALC7
photosphere. 
Its core intensity does not sense the presence of the ALC7
chromosphere but rather responds to disturbances in the low
photosphere.

The \NaIDone\ $S/B$ scattering ratio translates into a $b_u/b_l$ ratio
at left.
Their divergences would be the same in logarithmic units if $B$ and
$S$ were plotted directly, but I use representative temperatures to
remove nonlinear Wien sensitivities for comparisons with other
wavelengths.

The \NaIDone\ $b_l$ curve shows a steep increase in the chromosphere
because there Na ionization does not follow temperature (below).
This increase offsets the steep decline that the ground level
population would have according to the Saha equation (dotted). 

\parrr{Bound-free scattering}
In principle \acp{bf} scattering is the same as \acp{bb} scattering:
the equation description can be unified 
(\RTSAp{68}{Sect.~3.2.3} on page~48 of \RTSA). 
However, there are qualitative differences in important properties:

\begin{enumerate}

\item there is no coherency (\acp{PRD}) in \acp{bf} scattering because
each recombination employs a fresh electron; complete frequency
redistribution (\acp{CRD}) occurs over the full \acp{bf} edge
requiring averaging integrals over its extent (\RTSAp{93}{Eq.~3.109}
on page~73 of \RTSA);

\item the $\Lambda$ operator which produces angle-averaged radiation
intensity $J$ from the total source function (\RTSAp{101}{Sect.~4.1.3}
on page~81\,ff of \RTSA) gives $\Lambda(S) \tapprox S$ when
$S \sim 1+1.5\,\tau$, but produces $J > S$ for steeper $S(\tau)$
(Kourganoff's \RTSAp{103}{Fig.~4.4} on page~83 of \RTSA);

\item the $T(h)$ decay is set by radiative equilibrium $S \tapprox J$
in the optical part of the spectrum where the emergent flux peaks,
there forcing $S \sim 1+1.5\,\tau$ (\RTSAp{173}{Sect.~7.3.2} on
page~153\,ff of \RTSA).
In the ultraviolet Wien non-linearity produces steepening of the
Planck function for this given gradient (\RTSAp{121}{Fig.~4.9} on
page~101 of \RTSA);

\item a \acp{bf} edge has similar total (spectrum-integrated)
extinction as a \acp{bb} resonance line, but because it is much wider
its monochromatic extinction coefficient is much smaller. 
This means that the typical $B(\tau_\lambda)$ gradient remains
steeper, while in resonance lines it gets so shallow from the
additional line extinction that it approaches the isothermal case
obeying the $\sqrt\varepsilon$ law (\RTSAp{123}{Fig.~4.10}
on page~103 of \RTSA);

\item the solar continuum opacity does not only have its well-known
minimum at the \Hmin\,\acp{bf} threshold wavelength at 1.6~micron, but
also a (nearly as deep) minimum at 4000\,\AA\ because only shortward
of 4000\,\AA\ the metal \acp{bf} edges become more important than
\Hmin\,\acp{bf} extinction(Vitense's \RTSAp{199}{Fig.~8.6} on
page~179 of \RTSA).
Therefore, near-ultraviolet radiation escapes very deep and there the
$T(h)$ gradient is steep. 

\end{enumerate}

\begin{figure*}[t]
  \centerline{\includegraphics[width=0.99\textwidth]{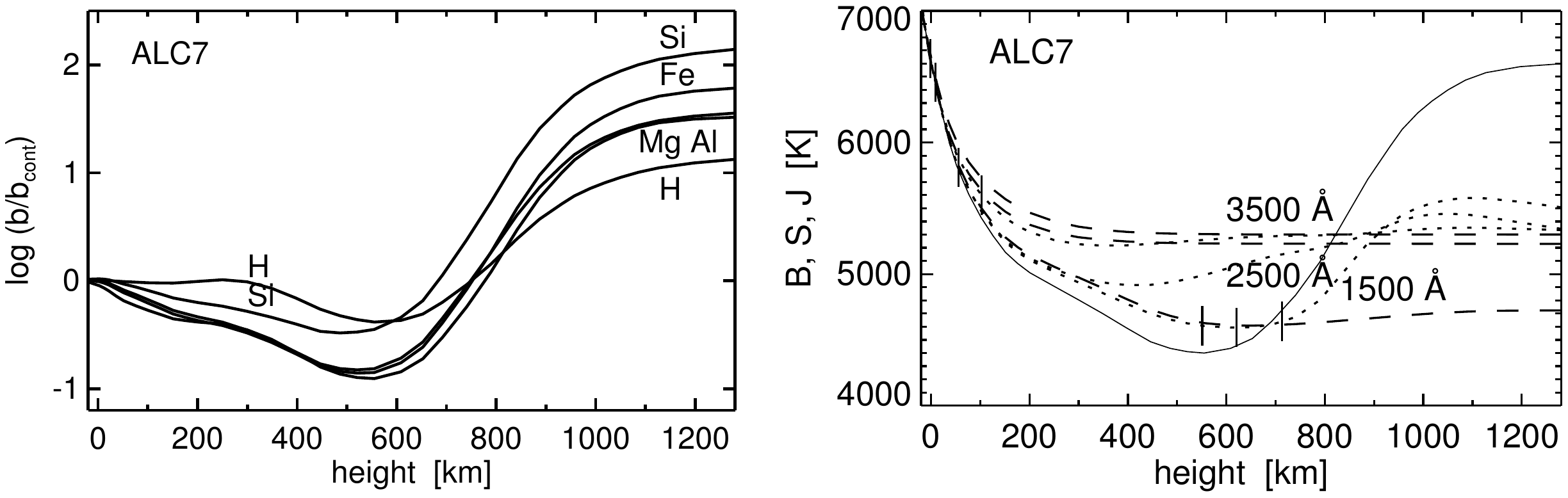}}
  \caption[]{\label{fig:uvedges}
  Left: departure coefficient ratios $b_2/b_{\rm cont}$ for \HI\ and
  $b_1/b_{\rm cont}$ for \MgI, \FeI, \SiI, \AlI.
  Right: $B$ (solid), $S$ (dotted) and $J$ (dashed) as representative
  temperatures at 3500, 2500, 1500\,\AA. 
  $S$ and $J$ are 100\,\AA\ averages including lines. 
  The ticks are at $\tau \tis 3, 1, 0.3$.
  }
\end{figure*}

The combined result of these \acp{bf} properties is that ultraviolet
continua escape by scattering out from the deep
photosphere\footnote{When the metal edges vanish through ionization
only the Balmer continuum remains with yet deeper escape.
This happens in kilogauss magnetic concentrations (``fluxtubes'') from
evacuation by magnetic pressure producing low gas density
(\citeads{1977PhDT.......237S}). 
Their brightening in 1600 and 1700\,\AA\ images from \acp{TRACE} and
\acp{AIA} is not from photospheric heating but from
hole-in-the-surface radiation just as ``line-gap'' brightening of
\FeI\ cores (\citeads{2009A&A...499..301V}). 
The larger brightening of Ellerman bursts at 1600 and 1700\,\AA\ is
also from photospheric metal ionization, but through intrinsic heating
at increased density (\citeads{2016A&A...590A.124R}).}, 
with $B$ gradients steeper than the resulting outward $S \tapprox J$
scattering declines. 
Figure~\ref{fig:uvedges} illustrates this for the Balmer continuum and
the electron-donor continua in the ultraviolet, whose edges
(wavelengths specified in
\linkadspage{1981ApJS...45..635V}{32}{Table~9} on page 665 of
\citeads{1981ApJS...45..635V} 
and in \RTSAp{196}{Table~8.1} on page~176 of \RTSA) together produce
the increasing extinction that makes the solar ultraviolet escape at
larger height for shorter wavelength. 
The $b$ ratios at left show a dip in the temperature minimum followed
by a steep increase that both result from following $J$ rather than
$B$, illustrated by the $S \tapprox J-B$ offsets at three ultraviolet
sample wavelengths at right. 

The electron donors have $b_{\rm cont} \tapprox 1$ because they are
largely ionized; their ratio curves at left represent under- and
overpopulation of their ground levels.
The deep initial dip for \FeI, MgI\ and \AlI\ implies that all
photospheric lines arising from their low-lying levels have increasing
and significant extinction depletion across the photosphere. 
For examples see my \linkcutssxp{10}{\MgI\,4571 from ALC7} and
\linkcutssxp{11}{\FeI\,6301.5 from ALC7}. 
Such depletion occurs generally for minority-species lines.

The exceptions are alkali lines which due to low ionization energy
suffer strong photon suction: population pull-down from the continuum
reservoir by photon losses in strong lines, see
\citetads{1992A&A...265..237B} 
or \RTSAp{233}{Sect.~10.1} on page~213\,ff of \RTSA.
It produces the initial hump rather than dip at 400\,km in the
\NaIDone\ $b_l$ curve in Fig.~\ref{fig:NaD}.

The scattering nature of ultraviolet continua has been illustrated 
best for the VALIIIC model (at 50 citations/year the most famous
plane-parallel star) in the wonderful eleven-page
\linkadspage{1981ApJS...45..635V}{33}{Fig.~36} on page 666\,ff of
\citetads{1981ApJS...45..635V}, 
with a selection copied on \RTSAp{204}{page~184\,ff} of \RTSA.

\parrr{Line haze modeling}
In the violet and ultraviolet myriads of neutral and once-ionized
metal lines together constitute a quasi-continuous ``line haze''
(\citeads{1980A&A....90..239G}) 
which lowers ultraviolet $J \!>\! B$ excesses.
Quantification is non-trivial because these lines are all scatterers
themselves, with much multi-level interlocking between multiplets and
multiplet members.
Their effect was still underestimated in the VALIII model; all Avrett
and Fontenla models since \citetads{1986ApJ...306..284M} 
therefore share a less steep photospheric temperature decline closer
to classic \acp{LTE} and radiative-equilibrium models
(\RTSAp{169}{Fig.~7.3} on page~149 of \RTSA).

\begin{figure*}[t]
  \centerline{\includegraphics[width=0.99\textwidth]{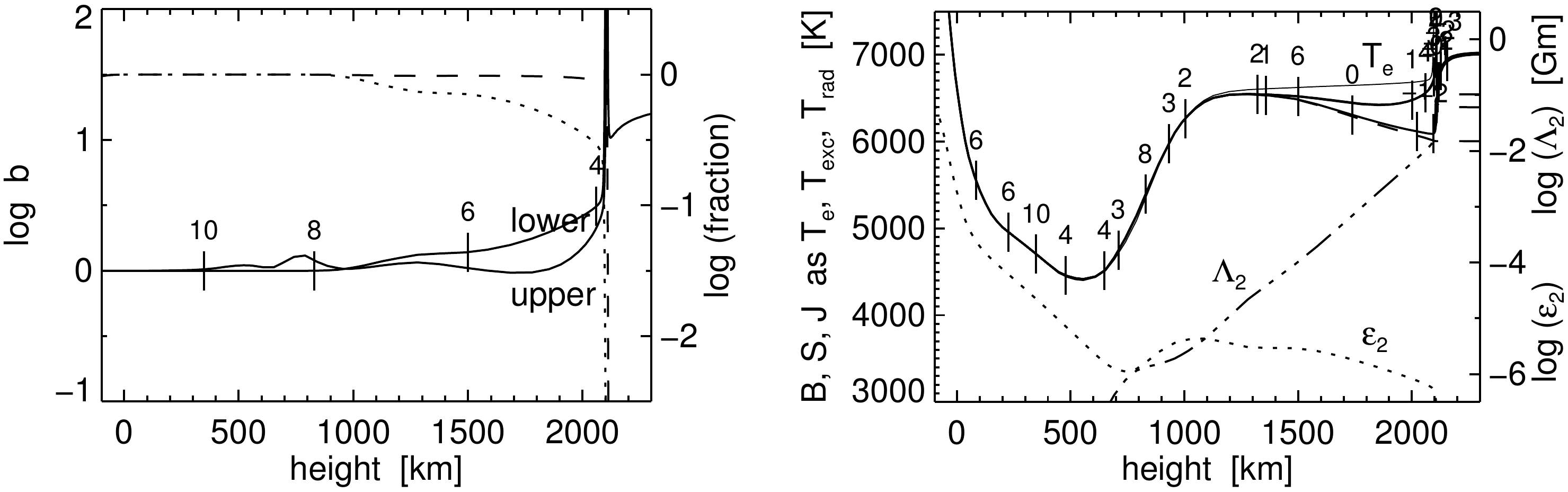}}
  \caption[]{\label{fig:Lya}
  \Lyalpha\ in the ALC7 chromosphere, in the format of
  Fig.~\ref{fig:NaD}. 
  Because \Lyalpha\ is a \acp{PRD} line there are three line source
  functions shown at right, for line center, the profile peaks, and
  the profile dips, each with (confusingly mixed) $\log \tau$ ticks
  but only the line-center ones at left. 
  }
\end{figure*}

\begin{figure*}[t]
  \centerline{\includegraphics[width=0.99\textwidth]{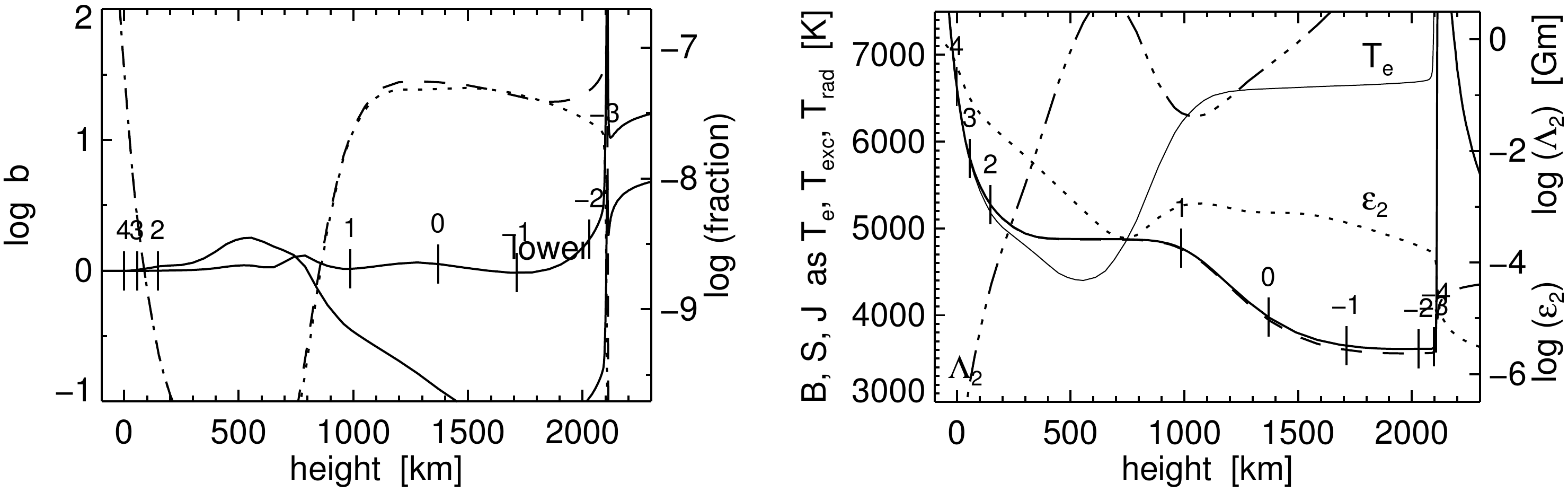}}
  \caption[]{\label{fig:Ha}
  \Halpha\ in and from the ALC7 chromosphere, in the format of
  Fig.~\ref{fig:NaD}. 
  }
\end{figure*}

In Figs.~\ref{fig:hydmodels}--\ref{fig:Ha} the line-haze lines are
accounted for with an individual two-level scattering approximation
(only the $\varepsilon$ terms in Eq.~\ref{eq:Sl}) 
in the RH code of \citetads{2001ApJ...557..389U} 
used for their production (with \acp{IDL} plotting programs available
on \href{https://www.staff.science.uu.nl/~rutte101/Recipes_IDL.html}
{my website}).
I used an RH setup sampling 340\,000 ultraviolet and optical lines
from the list of \citetads{2009AIPC.1171...43K} 
at 20\,m\AA\ wavelength spacing.
This approach overestimates line depths for many weaker lines (all
those connected to stronger ones via their upper level or via levels
close to their upper level), in particular for most optical lines from
the photosphere, by overestimating their scattering and resulting
$S\!<\!B$ departure.
However, in the ultraviolet where the line haze is most important such
scattering is a better assumption than \acp{LTE} $S \tis B$ which
gives non-observed central reversals for all lines with
$\tau \tapprox 1$ in the model chromosphere. 
In the construction of the ALC7 model all Kurucz lines were set to
transit from $S \tis B$ in the photosphere to $S \tis J$ in the
chromosphere with a gradual change-over factor comparable to
$\varepsilon$ in Eq.~\ref{eq:Sl}, the same for all lines
(\linkadspage{2008ApJS..175..229A}{15}{page 15} of
\citeads{2008ApJS..175..229A}). 
Both techniques still err in assuming \acp{SB} opacities for the
Kurucz lines, contrary to the \acp{NLTE} overionization affecting \eg\
all \FeI\ lines.
The most detailed inclusion of these line-haze lines through proper
multi-level \acp{NLTE} synthesis is in
\citetads{2015ApJ...809..157F}. 

So far, best-fit codes (incorrectly called inversion codes) that
iteratively adjust data-fitting atmosphere stratifications for lines
such as \FeI\,6301.5\,\AA\ ignore opacity \acp{NLTE} from ultraviolet
\acp{bf} depletion and the accompanying line haze issue.
They represent automation of the best-fit modeling by
\citetads{1967ZA.....65..365H} 
which led to the much-used model of
\citetads{1974SoPh...39...19H}. 
It ignored ultraviolet opacity depletion
(\citeads{1982A&A...115..104R}) 
and, worse for abundance determination, granulation
(\citeads{2004A&A...417..751A}). 
While complete \acp{NLTE} line-haze synthesis remains too demanding in
data-fit iteration, recipes as the one defined in
\citetads{1992A&A...265..237B} 
and available in RH may serve as shortcut estimator.

\parrr{\Lyalpha\ within ALC7}
\Lyalpha\ is the strongest and most scattering line in the solar
spectrum. 
Figure~\ref{fig:Lya} shows its properties across the ALC7
chromosphere.
Its scattering decline occurs higher up, in the transition region out
of \acp{ALMA} reach. 
The value of $\varepsilon$ is exceedingly small: \Lyalpha\ photons are
typically scattered a million times before getting absorbed. 
Nevertheless, because the extinction coefficient is enormous
(Fig.~\ref{fig:SB}) the Doppler-core thermalization length in
geometrical units $\sqrt{\pi}/\alpha\,\varepsilon$ is also small: only
of order $10^{-5}$\,Gm $=$ 10\,km in the low ALC7 chromosphere.
This is an underestimate because so deep in the atmosphere
($\tau \tapprox 10^8$ at line center!) the extended damping wings (from
linear Stark broadening with the Holtsmark distribution) become
important and provide longer scattering paths requiring
\acp{PRD} evaluation.
Below I estimate them to reach up to a few hundred km.

Even while scattering tremendously the \Lyalpha\ radiation remains
well contained locally within the ALC7 chromosphere.
It is boxed-in well enough that below the ALC7 transition region
detailed radiative balance is a good approximation: just as many
photons go up as go down, so that the net radiative rate is near zero
(\linkadspage{1981ApJS...45..635V}{29}{page~662} of
\citeads{1981ApJS...45..635V}). 
This means that $(b_l/b_u)\,J \tapprox B$, which holds when
$S \approx J$ without local radiative cooling or heating contribution
$\alpha\,(J-S)$ (\RTSAp{69}{page~49} of \RTSA).

Additionally, the short thermalization lengths thanks to the large
extinction cause also local collisional balancing in the \Lyalpha\
jump setting $J \approx B$, $b_2 \tapprox b_1$ and $S \tapprox B$ up to
large height (\RTSAp{71}{Eq.~3.41} on page~51 of \RTSA).
Since $b_1 \tapprox 1$ because the ground level contains almost all
hydrogen also $b_2 \tapprox 1$, as demonstrated in the lefthand panel
of Fig.~\ref{fig:Lya}. 

The \acp{LTE} fractional population curve (left, dotted) shows a
downturn from unity produced by increasing \acp{SB} ionization, but this
is compensated by a rise in $b_l$ which results from the onset of
\Lyalpha\ photon losses towards the transition region, especially in
the inner wings as shown by the multiple source functions at right. 
These have corresponding $b_2 \!<\! b_1$ divergence, but the $b_1$
rise compensates so that $b_2 \tapprox 1$ to large height.

Thus, the most scattering line in the solar spectrum has
near-\acp{LTE} extinction and a near-\acp{LTE} source function in the
ALC7 chromosphere.
The reasons are that the opacity in \Lyalpha\ is enormous from the
hydrogen abundance and from being the ground-level resonance line
without much ionization, and that detour paths via other hydrogen
transitions do not count.
\Lyalpha\ is the quintessential two-level scattering line, scatters
as it likes in detailed radiative balance, and so dictates the $n_2$
population governing other hydrogen rates.
The result is that within the ALC7 chromosphere \Lyalpha\ is an
\acp{LTE} agent in defining \Halpha\ extinction and constraining mm
extinction as discussed below. 
In the actual solar chromosphere it is also the major agent for these
but in \acp{non-E} fashion (below).

\parrr{\Halpha\ from ALC7}
Figure~\ref{fig:Ha} shows that \Halpha\ escapes from the middle ALC7
chromosphere.  
Its extinction obeys the \acp{SB} $\equiv$ \acp{LTE} estimate closely,
thanks to \Lyalpha. 
The fractional lower-level population therefore has a deep Boltzmann
dip in the temperature minimum which implies that one cannot observe
the upper photosphere in this line
(\citeads{1972SoPh...22..344S}). 
Indeed it primarily shows fibrils at line center and granules (and
occasional Ellerman bursts) in its wings but no shock scenes as in
\CaIIHK\
(\citeads{2008SoPh..251..533R}). 

Above the photosphere the thermalization length is above 10\,000\,km
so there is no chromospheric $S-B$ coupling.
Like \Lyalpha, \Halpha\ is pretty much its own boss in its scattering
and so sets $b_3$ with respect to $b_2$. 
\Lybeta\ shares $b_3$ and the \Halpha\ photon losses without affecting
these because it remains in detailed radiative balance at these
heights: when \Lybeta\ photons get converted into escaping \Halpha\
photons plus \Lyalpha\ photons then its $S \tapprox J$ diminishes
accordingly (Fig.~\ref{fig:hydmodels}).

The \Halpha\ source function shows a normal scattering decline across
the ALC7 chromosphere ($b_u \!<\! b_l$ divergence at left) that is
similar to that of \CaIR\ (\linkcutssxp{15}{\CaIR\ from ALC7}), but
it shows discordant behavior across the temperature minimum which is
due to radiation build-up across this opacity chasm by back-scattering
from the optically thick ALC7 chromosphere
(\citeads{2012A&A...540A..86R}). 
For \Halpha\ the ALC7 chromosphere is simply an overlying scattering
attenuator. 
Without it the scattering decline $S(\tau)$ would be the same but lie
in the photosphere as in Fig.~\ref{fig:NaD}; indeed a
radiative-equilibrium model without chromosphere has that and yet
produces the same \Halpha\ profile (``Ha-Ha'' comparison in
\linkadspage{2012A&A...540A..86R}{6}{Fig.~7} and
\linkadspage{2012A&A...540A..86R}{7}{Fig.~8} of
\citeads{2012A&A...540A..86R}). 

Most \Halpha\ photons that escape near 1400\,km from ALC7 were created
at the bottom of the photosphere where $\varepsilon$ is large.
They scatter out with increasing steps ($\Lambda$ curve), then
hit and scatter through the chromosphere until they escape around
$\tau \tis 1$. 
The actual large-contrast intensity pattern of the granulation imposed
at the creation height gets erased by the subsequent scattering, so
that lower-contrast chromospheric patterns (as fibrils) can yet show
up, as beautifully demonstrated in
\linkadspage{2012ApJ...749..136L}{7}{Fig.~7} of
\citetads{2012ApJ...749..136L} 
by comparing full 3D-scattering \Halpha\ synthesis with 1D column-wise
\Halpha\ synthesis.

The older \Halpha\ literature following
\citetads{1957ApJ...125..260T} 
and
\citetads{1959ApJ...129..401J} 
has called \Halpha\ photoelectrically-controlled, meaning domination
of the $\eta$ detour terms in Eq.~\ref{eq:Sl}, but in the ALC7 and
FCHHT-B atmospheres these contribute only 1\% to $S-J$ in the
temperature minimum and then rise to only about 5\% across the
chromosphere (\linkadspage{2012A&A...540A..86R}{9}{Fig.~12} of
\citeads{2012A&A...540A..86R}), 
as part of the ionization circuit discussed below.

The upshot is that \Halpha\ is a rather ordinary scatterer, be it with
special irradiation from below across its deep Boltzmann gap and with
special \acp{SB} opacity thanks to sitting on top of \Lyalpha. 
The emergent intensity $I \tapprox S(\tau\tis1)$ is lower for features
with larger opacity (\citeads{1972SoPh...22..344S}). 
In static models the detour contribution becomes dominant only in the
transition region; in the actual Sun when and where \Halpha\ gets very
bright from hydrogen recombination as in moss and flare rims in less
quiescent conditions.

\parrr{Hydrogen ionization in ALC7}
The $n_2$ population is fixed by \Lyalpha\ through detailed radiative
balancing and tight enclosure.
The top of the hydrogen atom from $n \tis 2$ at 10.2~eV to the
continuum at 13.6\,eV may be regarded as an extreme alkali-like atom
with only 3.4~eV ionization energy in which the Balmer lines act as
quasi-resonance lines and the quasi-ground-level population $n_2$ is
set by \Lyalpha.
The main processes are photo-ionization in the Balmer and higher
continua and cascade recombination along high levels.

In \acp{SE} a closed hydrogen-top circuit operates in which \acp{NLTE}
photo-ionization driving by
$T^{\rm rad}_{\rm BaC} \approx 5300\,\rmK \neq T_\rme$ (bottom panels
of Fig.~\ref{fig:hydmodels}) is balanced by cascade return along
levels $i \!>\! 4$ (\linkadspage{1981ApJS...45..635V}{30}{page~663} of
\citeads{1981ApJS...45..635V} 
but beware of the Menzel/Zwaan flip). 
The resulting $b_{\rm cont}$ curves represent additions to the
\Lyalpha-set $b_2$ curves, positive in the temperature minimum where
$T \!<\! 5300$\,K and negative in the chromosphere where
$T \!>\! 5300$\,K, as illustrated by its reverse $b_2/b_{\rm cont}$ in
Fig.~\ref{fig:uvedges}.

Although the large chromospheric \Halpha\ $S/B \tapprox b_3/b_2$
divergence is set by scattering photon losses in this line, these are
not an important driver in this circuit because their loss does not
increase $n_2$ as in alkali photon suction, due to the \Lyalpha-enforced 
$b_2 \approx b_1$ equilibrium with $n_1$ which is a much larger
electron reservoir than the continuum.
Let me clarify: when a \NaIDone\ photon leaves the Sun this enhances
the \NaI\ photo-ionization rate by putting an extra electron in the
\NaI\ ground level.
The \NaI\ line photon losses are a driver in establishing an
ionization/recombination replenishment circuit in \acp{SE}.
However, when an \Halpha\ photon leaves the Sun it leaves an electron
in level 2 that jumps down and back up a million or so times in
\Lyalpha\ before ending up in level 1 through collisional
deexcitation, unnoticeably enhancing the temperature and the gigantic
level-1 population.   
This way \Halpha\ is unusual in not increasing its lower-level $b$
with its photon losses as other strong lines do (\eg\
\linkcutssxp{15}{\CaIR\ from ALC7} and \linkcutssxp{16}{\CaIIK\ from
ALC7}) -- but those do not ride on \Lyalpha.

Overionization in the Balmer continuum is therefore the main
hydrogen-top driver.
The cascade recombination back to $n \tis 2$ preferentially follows
$\Delta n \tis -1$ steps downward, ending with \Halpha\ and making
that the main circuit-flow balancer to obtain \acp{SE}. 
Chromospheric radiative $J \!<\! S$ cooling in \Halpha\ therefore
roughly compensates radiative $J \!>\! S$ heating in the Balmer
continuum; both are much larger than such contributions by other
hydrogen transitions (\linkadspage{1981ApJS...45..635V}{69}{Fig.~48}
on page~68 of \citeads{1981ApJS...45..635V}, 
reproduced as \RTSAp{207}{Fig.~8.12} on page~187 of \RTSA). 

In the upper photosphere the hydrogen-top ionization-recombination
circuit produces emission in infrared \HI\ Rydberg lines
(\citeads{1992A&A...259L..53C}) 
similarly to the \MgI\ 12-micron emission lines
(\citeads{1992A&A...253..567C}; 
explanation in
\citeads{1994IAUS..154..309R}). 

\parrr{Hydrogen in ALC7 and FCHHT-B}
I now return to comparing hydrogen properties between these models
using Fig.~\ref{fig:hydmodels}.
In FCHHT-B the $S \tapprox J$ curves for \Lyalpha\ and \Lybeta\ have
extended tails below the steep onsets to the chromosphere and the
transition region that are not present in ALC7. 
They result from scattering over larger extent than would box-in the
Lyman-line radiation sufficiently to follow these abrupt temperature
changes. 
In the tails detailed radiative Lyman balancing still holds closely
($S \tapprox J \tapprox (b_u/b_l)\,B$) but detailed collisional
balancing does not hold ($b_u \neq b_l$).

The $S \tapprox J$ tails across the FCHHT-B temperature minimum come from
\Lyalpha\ and \Lybeta\ photons from the relatively hot chromosphere
penetrating a few hundred km below it.
The $S \tapprox J$ tails below the transition region penetrate over
similar extent. 
The transition region in ALC7 is as steep, but occurs at lower density
and has less effect.
The \Lyalpha\ tails cause increases of $b_2$ and increase \Halpha\
extinction substantially.

\Halpha\ behaves similarly in the two models apart from these boosts.
The larger optical thickness of the ALC7 chromosphere from higher
temperature and therefore Boltzmann population results in higher $J$
in the gap.
FCHHT-B has larger \Halpha\ opacity above 1500\,km but less lower down
(fractional population curves), resulting in slower $\tau$ buildup and
wider \Halpha\ core formation.

\Lybeta\ shares $b_3$ with \Halpha\ so where $b_2 \tapprox b_1$ they
have the same $S \tapprox (b_u/b_l) \, B$ behavior, but the
temperature representation separates their $S \tapprox J$ curves in
the bottom panels, with given $b_l/b_u$ giving larger
$T_\rme/T_{\rm exc}$ at longer wavelength.

The Balmer edge behaves the same in the two models because it is is
entirely defined by escape in the deep photosphere (Fig.~\ref{fig:uvedges}). 

\section{Hydrogen in dynamic non-E solar simulations}    
\label{sec:dynamic} \noindent
Non-E stands for non-\acp{LTE} plus non-\acp{SE}, \ie\ time-dependent
populations so that the rate equations do not sum to zero as in
\RTSAp{52}{Eq.~2.100} on page~32 of \RTSA.
For hydrogen it is usually called non-equilibrium ionization, but a
better name is non-equilibrium Lyman balancing because the large size
(10.2~eV) of the \Lyalpha\ jump is the main culprit.
It inhibits fast settling into detailed collisional \Lyalpha\
balancing at low temperature. 
The other hydrogen populations including $n_{\rm cont}$ follow suit.

The reason is the Boltzmann factor in
$C_{ul}/C_{lu} \tis (g_l/g_u) \exp(E_{ul}/kT)$ (\RTSAp{43}{Eq.~2.61}
on page~23 of \RTSA).
The downward collision rate $n_u\,C_{ul}$ barely depends on
temperature since any collider will do, regardless of speed.
Therefore, the upward \Lyalpha\ collision rate $n_1\,C_{12}$ shares
the very steep Boltzmann increase shown by the \Halpha\ curve in
Fig.~\ref{fig:SB}.
Both rates scale with the electron density (\RTSAp{70}{Eqs.~3.32-3.33}
on page~50 of \RTSA).

The pertinent \acp{non-E} publications describe simulations with the
1D HD code RADYN
(\citeads{2002ApJ...572..626C}), 
the 2D MHD code Stagger
(\citeads{2007A&A...473..625L}), 
and the 3D MHD code Bifrost
(\citeads{2012ApJ...749..136L}; 
\citeads{2015A&A...575A..15L}; 
\citeads{2016A&A...585A...4C}). 
In each simulation ubiquitous shocks occur that reach temperatures
around 7000\,K at electron densities around $10^{11}$\,cm$^{-3}$.
The RADYN simulation showed that at chromospheric heights the Lyman
lines remain close to detailed radiative balance in and after these
shocks; this condition was then set as tractability assumption in the
more demanding multi-D simulations.

Shocks are hot and dense and also have increased $N_\rme/N_\rmH$ from
partial hydrogen ionization. 
Hence in shocks $C_{12}$ is large and collisional \Lyalpha\ balancing
is fast, reaching equilibrium within seconds. 
On the contrary, in shock aftermaths in which the gas cools to low
temperature and all hydrogen recombines, the settling time scale
becomes very long, $10^3$~seconds or more
(\linkadspage{2002ApJ...572..626C}{7}{Fig.~6} of
\citeads{2002ApJ...572..626C}). 

In the shocks \Lyalpha\ reaches $S \tapprox B$ just as within the ALC7
chromosphere (which has similar temperature and electron density). 
This was demonstrated in the bottom panels of
\linkadspage{2007A&A...473..625L}{5}{Fig.~2}
(\href{http://www.staff.science.uu.nl/~rutte101/rrweb/rjr-talkstuff/hion2_fig2_movie.mov}{movie
version}) of \citetads{2007A&A...473..625L} 
where the \acp{non-E} $n_2$ curves equal the $\acp{LTE}$ curves
momentarily in shocks.
However, the same figure shows that in the cooling post-shock
aftermaths the hydrogen $n_2$ population hangs at the large \acp{LTE}
shock value until the next shock arrives, typically 3--5 minutes
later, whereas the \acp{LTE} prediction follows the actual temperature
instantaneously and drops dramatically following the Boltzmann slopes
in Fig.~\ref{fig:SB}.
The last panel of \linkadspage{2007A&A...473..625L}{4}{Fig.~1}
(\href{http://www.staff.science.uu.nl/~rutte101/rrweb/rjr-talkstuff/hion2_fig1_movie.mov}{movie
version}) of \citetads{2007A&A...473..625L} shows that $b_2$ can so
reach values above 12~dex! 
While these are called \acp{NLTE} overpopulations they actually 
are \acp{LTE} underpopulations because that assumption is the
bad one.

The hydrogen ionization is again set by balancing Balmer
photo-ionization and cascade recombination
(\linkadspage{2002ApJ...572..626C}{4}{Fig.~3} of
\citeads{2002ApJ...572..626C}). 
This hydrogen-top loop balances fast and again adds
\acp{NLTE}-\acp{SE} $b_{\rm cont}$ additions to the $b_2$ value
controlled by \Lyalpha, positive or negative for temperature below or
above the Balmer continuum radiation temperature of 5300\,K. 
This is illustrated in the second-to-last panel of
\linkadspage{2007A&A...473..625L}{4}{Fig.~1} of
\citetads{2007A&A...473..625L} by the wider span of the $b_{\rm cont}$
color scale, reaching above 15~dex in 2400\,K aftermath clouds (the
lower limit imposed in the simulation, \cf\
\citeads{2011A&A...530A.124L}) 
in which the Balmer-continuum driving ratio
$B_{\rm BaC}(5300\,\rmK)/B_{\rm BaC}(2400\,\rmK)$ is 4~dex.

\section{Hydrogen in solar contrails}
\label{sec:contrails} \noindent
My premise in \PubI\ is that many if not most long \Halpha\ fibrils
represent contrails of cooling gas in which hydrogen recombines, each
outlining the track of a small fast propagating heating event in which
hydrogen got ionized that passed minutes before, as in the example of
\citetads{2016arXiv160907616R}. 

This suggestion implies sudden precursor heating, as in the simulation
shocks but to yet higher temperature and electron density and
therefore with faster \Lyalpha\ balancing giving \acp{SB} extinction
to \Halpha\ that becomes very large above 8000\,K (Fig.~\ref{fig:SB}). 
In the subsequent cooling and recombining aftermath the \Lyalpha\
balancing becomes slow at low temperature, maintaining the large
initial \Halpha\ extinction at growing $b_2$ overpopulation while
$n_2$ hangs at its high earlier \acp{SB} value, instead of
instantaneously following the diminishing temperature along the steep
Boltzmann slopes in Fig.~\ref{fig:SB} down to very small extinction.
I call this large \acp{non-E} post-hot extinction
``post-Saha-Boltzmann-extinction'' (PSBE) and suggest that it
furnishes \Halpha\ contrail visibility enriching observed solar
\Halpha\ scenes with long fibrils.

For the \HI\,\acp{ff} mm continua governing solar \acp{ALMA} images
the instantaneous \acp{SE} ionization/recombination loop in the
hydrogen top causes a tilt-like modification of the steep \Halpha\
Boltzmann slopes in Fig.~\ref{fig:SB}, with the Balmer
continuum radiation temperature of 5300\,K as pivot: diminishing the
extinction from the \acp{SB} value for $T \!>\! 5300$\,K and
increasing it for $T \!<\! 5300$\,K.
However, such tilt corrections amount to only a few dex, minor
compared with the 10-dex Boltzmann increase.
At near-complete hydrogen ionization the \HI\,\acp{ff} mm extinction
anyhow remains as large as shown in Fig.~\ref{fig:SB}, at
longer wavelengths very much larger than the \Halpha\ extinction.
Therefore I predict large mm extinction, hence \acp{ALMA} visibility,
for any feature with \acp{PSBE} contrail visibility in \Halpha.

Lateral contrail contrasts then depend on contrail histories.
\acp{ALMA} observes the current temperature from \acp{PSBE}-boosted
optically thick features, but for \Halpha\ the history not only boosts 
the feature opacity but larger opacity also darkens the line core
along the scattering decline.
\Halpha\ fibrils may so differ in darkness at similar temperature
while the same fibrils seen with \acp{ALMA} do not.
Also, unlike \Halpha\ \acp{ALMA} does not sense difference in 
Dopplershift.
I therefore expect contrails to appear less fibrilar and more
blanket-like in \acp{ALMA} images.

The \Lyalpha\ back-radiation tails for FCHHT-B in
Fig.~\ref{fig:hydmodels} suggest that \Lyalpha\ scattering from
precursors and contrails into cool surroundings produces aureole
visibility by increasing the $n_2$ population and hence the
\HI\,\acp{ff} mm extinction.

\section{Discussion} \label{sec:discussion} \noindent
\parrr{Summary}
Chromospheric hydrogen diagnostics are dominated by the 10.2~eV
\Lyalpha\ jump. 
Its large size causes very steep initial temperature sensitivity for
\Halpha\ and \HI\,\acp{ff} extinction (Fig.~\ref{fig:SB}) and long
recombination retardation during post-ionization cooling which implies
large-extinction aftermaths of heating events in \Halpha\ and 
\acp{ALMA} images.

\parrr{Precursors}
The nature of the proposed heating events producing long \Halpha\
fibrils as \acp{PSBE} aftermath contrails is not known.
The precursor in the example of
\citetads{2016arXiv160907616R} 
appeared rather like spicules-II of which the agent is also unknown
(\citeads{2012ApJ...759...18P}), 
but with more horizontal launching.
So far spicules-II and long \Halpha\ fibrils lack in Bifrost
simulations and therefore also in the \acp{ALMA} predictions of
\citetads{2015A&A...575A..15L}. 
Bipolar ion-neutral separation may play a role where hydrogen ionizes
(\citeads{2016ApJ...831L...1M}). 
Small-scale torsion kicking may deliver long fibril extent because
torsion waves are incompressive, travel far before dissipation, and
appear ubiquitously
(\citeads{2014Sci...346D.315D}). 

\acp{ALMA} may be the best facility to study contrail precursors
thanks to their large \HI\,\acp{ff} extinction.
When \acp{ALMA} reaches sufficient angular resolution it may
faithfully map precursor morphology since free-free continua do not
suffer scattering. 
It may also map cooler precursor and contrail aureoles that gain
opacity from \Lyalpha\ surround scattering.

\parrr{Rydberg lines}
A final solar-\acp{ALMA} item (not in \PubI\ as too speculative for
prediction) concerns high-$n$ \HI\ lines. 
I suggest that some may be detectable with \acp{ALMA}.

In the infrared the \MgI\ 7--6 lines at 12.2 and 12.3~micron due to
ladder-wise cascade recombination as a replenishment flow (explanation
in \citeads{1994IAUS..154..309R}) 
are more conspicuous than the similarly formed \HI\ 7--6 line at
12.4~micron because the \MgI\ $n \tis 6$ levels have larger population
in the upper photosphere
(\linkadspage{1992A&A...253..567C}{16}{Fig.~15} of
\citeads{1992A&A...253..567C}). 
Towards higher $n$ the different Rydberg wavelengths converge to the
hydrogenic limit so that higher-$n$ lines become blended mixtures
of different species.

The highest $n$ levels, up to the formal $n\tis\infty$, vanish into
the continuum due to collisional ionization-limit lowering; the limit
for existence depends on density (pages 69 and 275 of
\citeads{1970stat.book.....M}). 
However, the EK results in Figs.~4 and 5 of
\citetads{1992ApJ...396..364K} 
suggest that Rydberg levels with $\alpha$ ($\Delta n \tis 1$) lines
between them in the \acp{ALMA} range may exist in chromospheric
conditions.

\linkadspage{1992A&A...259L..53C}{4}{Figure~3} of
\citetads{1992A&A...259L..53C} 
shows predicted emission peak strengths $I/I_{\rm cont}$ for the \HI\
$\alpha$ lines with lower levels 4 to 18 for the plane-parallel
\acp{SE} model of \citetads{1986ApJ...306..284M} 
and suggests that these become negligible at higher $n$.
The highest detected so far are 19-$\alpha$
(\citeads{2000A&A...357..757C}) 
and 21-$\alpha$
(\citeads{2000A&A...361L..60C}), 
but only at the limb and fitting the predicted weakening
(\linkadspage{2000A&A...357..757C}{5}{Fig.~6} of
\citeads{2000A&A...357..757C}). 

However, my proposed mechanisms for long \Halpha\ fibrils give all
hydrogen-top levels much larger population than static-atmosphere
\acp{SE} modeling predicts.
If they are not swamped by the also boosted \HI\,\acp{ff} background
\HI\ Rydberg lines may be detectable in \acp{ALMA} spectra and be freed
of non-boosted blends. 
They may then appear as small and fast spectral emission features on the disk
arising from heating events with steep temperature gradients and
resulting $T(\tau_\mu \tis 1)$ contrasts, and as coarser
spectral-extent features at the limb due to \acp{PSBE}-boosted cooling
contrails with less temperature contrast but $\tau\,T$ visibility
against the dark-sky background.

In the coming solar observing during \acp{ALMA} Cycle~4 the
\HI\,30-$\alpha$ line at 231.901\,GHz sits near the center of the
2-GHz wide sampling interval around 232\,GHz of Band~6 in which
0.5~arcsec resolution can be reached with the 538.9~m baseline.
This represents the best candidate.
In addition, 39-$\alpha$ at 106.737\,GHz and 41-$\alpha$ at
92.034\,GHz fall possibly just inside such sampling intervals at 105
and 93\,GHz in Band~3
(\citeads{2016ASPC..504..327K}; 
Rydberg frequencies from the
\href{http://physics.nist.gov/PhysRefData/HDEL/transfreq.html}{NIST
calculator}).

If such lines are detected then measuring their
Zeeman broadening or merging
(\citeads{1975SoPh...44..371G}; 
\cf\ \citeads{1980A&A....82..388G}) 
or even partial splitting as in
\linkadspage{1994IAUS..154..309R}{12}{Fig.~9} of
\citetads{1994IAUS..154..309R} 
is an exciting prospect for magnetometry of the \acp{non-E} chromosphere.

\begin{small} 

\begin{acknowledgements} \noindent I thank S.~Toriumi for inviting me
to the National Astronomical Observatory of Japan where I made most of
the above figures, the organizers of \acp{IAU} Symposium 327 for a
worthwhile and pleasant conference, and the Leids Kerkhoven-Bosscha Fonds
for travel support.
\end{acknowledgements}


\begin{thebibliography}{55}
\expandafter\ifx\csname natexlab\endcsname\relax\def\natexlab#1{#1}\fi

\bibitem[{{Asensio Ramos} {et~al.}(2003){Asensio Ramos}, {Trujillo Bueno},
  {Carlsson}, \& {Cernicharo}}]{2003ApJ...588L..61A}
{Asensio Ramos}, A., {Trujillo Bueno}, J., {Carlsson}, M., \& {Cernicharo}, J.
  2003, \apjl, 588, L61 \csname 2003ApJ...588L..61Alink\endcsname~\csname
  2003ApJ...588L..61Anote\endcsname

\bibitem[{{Asplund} {et~al.}(2004){Asplund}, {Grevesse}, {Sauval}, {Allende
  Prieto}, \& {Kiselman}}]{2004A&A...417..751A}
{Asplund}, M., {Grevesse}, N., {Sauval}, A.~J., {Allende Prieto}, C., \&
  {Kiselman}, D. 2004, \aap, 417, 751 \csname
  2004A&A...417..751Alink\endcsname~\csname 2004A&A...417..751Anote\endcsname

\bibitem[{{Avrett}(1965)}]{1965SAOSR.174..101A}
{Avrett}, E.~H. 1965, SAO Special Report, 174, 101 \csname
  1965SAOSR.174..101Alink\endcsname~\csname 1965SAOSR.174..101Anote\endcsname

\bibitem[{{Avrett} \& {Loeser}(2008)}]{2008ApJS..175..229A}
{Avrett}, E.~H. \& {Loeser}, R. 2008, \apjs, 175, 229 \csname
  2008ApJS..175..229Alink\endcsname~\csname 2008ApJS..175..229Anote\endcsname

\bibitem[{{Bruls} {et~al.}(1992){Bruls}, {Rutten}, \&
  {Shchukina}}]{1992A&A...265..237B}
{Bruls}, J.~H.~M.~J., {Rutten}, R.~J., \& {Shchukina}, N.~G. 1992, \aap, 265,
  237 \csname 1992A&A...265..237Blink\endcsname~\csname
  1992A&A...265..237Bnote\endcsname

\bibitem[{{Carlsson} {et~al.}(2016){Carlsson}, {Hansteen}, {Gudiksen},
  {Leenaarts}, \& {De Pontieu}}]{2016A&A...585A...4C}
{Carlsson}, M., {Hansteen}, V.~H., {Gudiksen}, B.~V., {Leenaarts}, J., \& {De
  Pontieu}, B. 2016, \aap, 585, A4 \csname
  2016A&A...585A...4Clink\endcsname~\csname 2016A&A...585A...4Cnote\endcsname

\bibitem[{{Carlsson} \& {Rutten}(1992)}]{1992A&A...259L..53C}
{Carlsson}, M. \& {Rutten}, R.~J. 1992, \aap, 259, L53 \csname
  1992A&A...259L..53Clink\endcsname~\csname 1992A&A...259L..53Cnote\endcsname

\bibitem[{{Carlsson} {et~al.}(1992){Carlsson}, {Rutten}, \&
  {Shchukina}}]{1992A&A...253..567C}
{Carlsson}, M., {Rutten}, R.~J., \& {Shchukina}, N.~G. 1992, \aap, 253, 567
  \csname 1992A&A...253..567Clink\endcsname~\csname
  1992A&A...253..567Cnote\endcsname

\bibitem[{{Carlsson} \& {Stein}(1995)}]{1995ApJ...440L..29C}
{Carlsson}, M. \& {Stein}, R.~F. 1995, \apjl, 440, L29 \csname
  1995ApJ...440L..29Clink\endcsname~\csname 1995ApJ...440L..29Cnote\endcsname

\bibitem[{{Carlsson} \& {Stein}(2002)}]{2002ApJ...572..626C}
{Carlsson}, M. \& {Stein}, R.~F. 2002, \apj, 572, 626 \csname
  2002ApJ...572..626Clink\endcsname~\csname 2002ApJ...572..626Cnote\endcsname

\bibitem[{{Clark} {et~al.}(2000{\natexlab{a}}){Clark}, {Naylor}, \&
  {Davis}}]{2000A&A...357..757C}
{Clark}, T.~A., {Naylor}, D.~A., \& {Davis}, G.~R. 2000{\natexlab{a}}, \aap,
  357, 757 \csname 2000A&A...357..757Clink\endcsname~\csname
  2000A&A...357..757Cnote\endcsname

\bibitem[{{Clark} {et~al.}(2000{\natexlab{b}}){Clark}, {Naylor}, \&
  {Davis}}]{2000A&A...361L..60C}
{Clark}, T.~A., {Naylor}, D.~A., \& {Davis}, G.~R. 2000{\natexlab{b}}, \aap,
  361, L60 \csname 2000A&A...361L..60Clink\endcsname~\csname
  2000A&A...361L..60Cnote\endcsname

\bibitem[{{De Pontieu} {et~al.}(2014){De Pontieu}, {Rouppe van der Voort},
  {McIntosh}, {Pereira}, {Carlsson}, {Hansteen}, {Skogsrud}, {Lemen}, {Title},
  {Boerner}, {Hurlburt}, {Tarbell}, {Wuelser}, {De Luca}, {Golub}, {McKillop},
  {Reeves}, {Saar}, {Testa}, {Tian}, {Kankelborg}, {Jaeggli}, {Kleint}, \&
  {Mart{\'{\i}}nez-Sykora}}]{2014Sci...346D.315D}
{De Pontieu}, B., {Rouppe van der Voort}, L., {McIntosh}, S.~W., {et~al.} 2014,
  Science, 346, 1255732 \csname 2014Sci...346D.315Dlink\endcsname~\csname
  2014Sci...346D.315Dnote\endcsname

\bibitem[{{Ellerman}(1917)}]{1917ApJ....46..298E}
{Ellerman}, F. 1917, \apj, 46, 298 \csname
  1917ApJ....46..298Elink\endcsname~\csname 1917ApJ....46..298Enote\endcsname

\bibitem[{{Fontenla} {et~al.}(1993){Fontenla}, {Avrett}, \&
  {Loeser}}]{1993ApJ...406..319F}
{Fontenla}, J.~M., {Avrett}, E.~H., \& {Loeser}, R. 1993, \apj, 406, 319
  \csname 1993ApJ...406..319Flink\endcsname~\csname
  1993ApJ...406..319Fnote\endcsname

\bibitem[{{Fontenla} {et~al.}(2009){Fontenla}, {Curdt}, {Haberreiter},
  {Harder}, \& {Tian}}]{2009ApJ...707..482F}
{Fontenla}, J.~M., {Curdt}, W., {Haberreiter}, M., {Harder}, J., \& {Tian}, H.
  2009, \apj, 707, 482 \csname 2009ApJ...707..482Flink\endcsname~\csname
  2009ApJ...707..482Fnote\endcsname

\bibitem[{{Fontenla} {et~al.}(2015){Fontenla}, {Stancil}, \&
  {Landi}}]{2015ApJ...809..157F}
{Fontenla}, J.~M., {Stancil}, P.~C., \& {Landi}, E. 2015, \apj, 809, 157
  \csname 2015ApJ...809..157Flink\endcsname~\csname
  2015ApJ...809..157Fnote\endcsname

\bibitem[{{Greve}(1975)}]{1975SoPh...44..371G}
{Greve}, A. 1975, \solphys, 44, 371 \csname
  1975SoPh...44..371Glink\endcsname~\csname 1975SoPh...44..371Gnote\endcsname

\bibitem[{{Greve} \& {Pauls}(1980)}]{1980A&A....82..388G}
{Greve}, A. \& {Pauls}, T. 1980, \aap, 82, 388 \csname
  1980A&A....82..388Glink\endcsname~\csname 1980A&A....82..388Gnote\endcsname

\bibitem[{{Greve} \& {Zwaan}(1980)}]{1980A&A....90..239G}
{Greve}, A. \& {Zwaan}, C. 1980, \aap, 90, 239 \csname
  1980A&A....90..239Glink\endcsname~\csname 1980A&A....90..239Gnote\endcsname

\bibitem[{{Holweger}(1967)}]{1967ZA.....65..365H}
{Holweger}, H. 1967, \zap, 65, 365 \csname
  1967ZA.....65..365Hlink\endcsname~\csname 1967ZA.....65..365Hnote\endcsname

\bibitem[{{Holweger} \& {M{\"{u}}ller}(1974)}]{1974SoPh...39...19H}
{Holweger}, H. \& {M{\"{u}}ller}, E.~A. 1974, \solphys, 39, 19 \csname
  1974SoPh...39...19Hlink\endcsname~\csname 1974SoPh...39...19Hnote\endcsname

\bibitem[{{Jefferies}(1968)}]{1968slf..book.....J}
{Jefferies}, J.~T. 1968, {Spectral line formation} \csname
  1968slf..book.....Jlink\endcsname~\csname 1968slf..book.....Jnote\endcsname

\bibitem[{{Jefferies} \& {Thomas}(1959)}]{1959ApJ...129..401J}
{Jefferies}, J.~T. \& {Thomas}, R.~N. 1959, \apj, 129, 401 \csname
  1959ApJ...129..401Jlink\endcsname~\csname 1959ApJ...129..401Jnote\endcsname

\bibitem[{{Kobelski} \& {ALMA Solar Development
  Team}(2016)}]{2016ASPC..504..327K}
{Kobelski}, A. \& {ALMA Solar Development Team}. 2016, in Astron.\ Soc.\
  Pacific Conf.\ Series, Vol. 504, Coimbra Solar Physics Meeting: Ground-based
  Solar Observations in the Space Instrumentation Era, ed. I.~{Dorotovic},
  C.~E. {Fischer}, \& M.~{Temmer}, 327 \csname
  2016ASPC..504..327Klink\endcsname~\csname 2016ASPC..504..327Knote\endcsname

\bibitem[{{Kunc} \& {Soon}(1992)}]{1992ApJ...396..364K}
{Kunc}, J.~A. \& {Soon}, W.~H. 1992, \apj, 396, 364 \csname
  1992ApJ...396..364Klink\endcsname~\csname 1992ApJ...396..364Knote\endcsname

\bibitem[{{Kurucz}(2009)}]{2009AIPC.1171...43K}
{Kurucz}, R.~L. 2009, in Am.\ Inst. Phys.\ Conf.\ Series, ed.
  I.~{Huben{\'{y}}}, J.~M. {Stone}, K.~{MacGregor}, \& K.~{Werner}, Vol. 1171,
  43--51 \csname 2009AIPC.1171...43Klink\endcsname~\csname
  2009AIPC.1171...43Knote\endcsname

\bibitem[{{Leenaarts} {et~al.}(2011){Leenaarts}, {Carlsson}, {Hansteen}, \&
  {Gudiksen}}]{2011A&A...530A.124L}
{Leenaarts}, J., {Carlsson}, M., {Hansteen}, V., \& {Gudiksen}, B.~V. 2011,
  \aap, 530, A124 \csname 2011A&A...530A.124Llink\endcsname~\csname
  2011A&A...530A.124Lnote\endcsname

\bibitem[{{Leenaarts} {et~al.}(2007){Leenaarts}, {Carlsson}, {Hansteen}, \&
  {Rutten}}]{2007A&A...473..625L}
{Leenaarts}, J., {Carlsson}, M., {Hansteen}, V., \& {Rutten}, R.~J. 2007, \aap,
  473, 625 \csname 2007A&A...473..625Llink\endcsname~\csname
  2007A&A...473..625Lnote\endcsname

\bibitem[{{Leenaarts} {et~al.}(2012){Leenaarts}, {Carlsson}, \& {Rouppe van der
  Voort}}]{2012ApJ...749..136L}
{Leenaarts}, J., {Carlsson}, M., \& {Rouppe van der Voort}, L. 2012, \apj, 749,
  136 \csname 2012ApJ...749..136Llink\endcsname~\csname
  2012ApJ...749..136Lnote\endcsname

\bibitem[{{Loukitcheva} {et~al.}(2015){Loukitcheva}, {Solanki}, {Carlsson}, \&
  {White}}]{2015A&A...575A..15L}
{Loukitcheva}, M., {Solanki}, S.~K., {Carlsson}, M., \& {White}, S.~M. 2015,
  \aap, 575, A15 \csname 2015A&A...575A..15Llink\endcsname~\csname
  2015A&A...575A..15Lnote\endcsname

\bibitem[{{Maltby} {et~al.}(1986){Maltby}, {Avrett}, {Carlsson},
  {Kjeldseth-Moe}, {Kurucz}, \& {Loeser}}]{1986ApJ...306..284M}
{Maltby}, P., {Avrett}, E.~H., {Carlsson}, M., {et~al.} 1986, \apj, 306, 284
  \csname 1986ApJ...306..284Mlink\endcsname~\csname
  1986ApJ...306..284Mnote\endcsname

\bibitem[{{Mart{\'{\i}}nez-Sykora} {et~al.}(2016){Mart{\'{\i}}nez-Sykora}, {De
  Pontieu}, {Carlsson}, \& {Hansteen}}]{2016ApJ...831L...1M}
{Mart{\'{\i}}nez-Sykora}, J., {De Pontieu}, B., {Carlsson}, M., \& {Hansteen},
  V. 2016, \apjl, 831, L1 \csname 2016ApJ...831L...1Mlink\endcsname~\csname
  2016ApJ...831L...1Mnote\endcsname

\bibitem[{{Mihalas}(1970)}]{1970stat.book.....M}
{Mihalas}, D. 1970, {Stellar atmospheres} \csname
  1970stat.book.....Mlink\endcsname~\csname 1970stat.book.....Mnote\endcsname

\bibitem[{{Pereira} {et~al.}(2012){Pereira}, {De Pontieu}, \&
  {Carlsson}}]{2012ApJ...759...18P}
{Pereira}, T.~M.~D., {De Pontieu}, B., \& {Carlsson}, M. 2012, \apj, 759, 18
  \csname 2012ApJ...759...18Plink\endcsname~\csname
  2012ApJ...759...18Pnote\endcsname

\bibitem[{{Rutten}(1988)}]{1988ASSL..138..185R}
{Rutten}, R.~J. 1988, in Astrophys.\ Space Sci.\ Library, Vol. 138, IAU
  Colloq.\ 94: Physics of Formation of Fe II Lines Outside LTE, ed.
  R.~{Viotti}, A.~{Vittone}, \& M.~{Friedjung}, 185 \csname
  1988ASSL..138..185Rlink\endcsname~\csname 1988ASSL..138..185Rnote\endcsname

\bibitem[{{Rutten}(2002)}]{2002JAD.....8....8R}
{Rutten}, R.~J. 2002, Journal of Astronomical Data, 8 \csname
  2002JAD.....8....8Rlink\endcsname~\csname 2002JAD.....8....8Rnote\endcsname

\bibitem[{{Rutten}(2003)}]{2003rtsa.book.....R}
{Rutten}, R.~J. 2003, {Radiative Transfer in Stellar Atmospheres} \csname
  2003rtsa.book.....Rlink\endcsname~\csname 2003rtsa.book.....Rnote\endcsname

\bibitem[{{Rutten}(2016{\natexlab{a}})}]{2016arXiv160901122R}
{Rutten}, R.~J. 2016{\natexlab{a}}, ArXiv e-prints \csname
  2016arXiv160901122Rlink\endcsname~\csname 2016arXiv160901122Rnote\endcsname

\bibitem[{{Rutten}(2016{\natexlab{b}})}]{2016A&A...590A.124R}
{Rutten}, R.~J. 2016{\natexlab{b}}, \aap, 590, A124 \csname
  2016A&A...590A.124Rlink\endcsname~\csname 2016A&A...590A.124Rnote\endcsname

\bibitem[{{Rutten} \& {Carlsson}(1994)}]{1994IAUS..154..309R}
{Rutten}, R.~J. \& {Carlsson}, M. 1994, in IAU Symposium, Vol. 154, Infrared
  Solar Physics, ed. D.~M. {Rabin}, J.~T. {Jefferies}, \& C.~{Lindsey}, 309
  \csname 1994IAUS..154..309Rlink\endcsname~\csname
  1994IAUS..154..309Rnote\endcsname

\bibitem[{{Rutten} \& {Kostik}(1982)}]{1982A&A...115..104R}
{Rutten}, R.~J. \& {Kostik}, R.~I. 1982, \aap, 115, 104 \csname
  1982A&A...115..104Rlink\endcsname~\csname 1982A&A...115..104Rnote\endcsname

\bibitem[{{Rutten} \& {Rouppe van der Voort}(2016)}]{2016arXiv160907616R}
{Rutten}, R.~J. \& {Rouppe van der Voort}, L.~H.~M. 2016, ArXiv e-prints
  \csname 2016arXiv160907616Rlink\endcsname~\csname
  2016arXiv160907616Rnote\endcsname

\bibitem[{{Rutten} \& {Uitenbroek}(2012)}]{2012A&A...540A..86R}
{Rutten}, R.~J. \& {Uitenbroek}, H. 2012, \aap, 540, A86 \csname
  2012A&A...540A..86Rlink\endcsname~\csname 2012A&A...540A..86Rnote\endcsname

\bibitem[{{Rutten} {et~al.}(2008){Rutten}, {van Veelen}, \&
  {S{\"u}tterlin}}]{2008SoPh..251..533R}
{Rutten}, R.~J., {van Veelen}, B., \& {S{\"u}tterlin}, P. 2008, \solphys, 251,
  533 \csname 2008SoPh..251..533Rlink\endcsname~\csname
  2008SoPh..251..533Rnote\endcsname

\bibitem[{{Rutten} {et~al.}(2013){Rutten}, {Vissers}, {Rouppe van der Voort},
  {S{\"u}tterlin}, \& {Vitas}}]{2013JPhCS.440a2007R}
{Rutten}, R.~J., {Vissers}, G.~J.~M., {Rouppe van der Voort}, L.~H.~M.,
  {S{\"u}tterlin}, P., \& {Vitas}, N. 2013, J.\ Phys.\ Conf.\ Ser., 440, 012007
  \csname 2013JPhCS.440a2007Rlink\endcsname~\csname
  2013JPhCS.440a2007Rnote\endcsname

\bibitem[{{Schoolman}(1972)}]{1972SoPh...22..344S}
{Schoolman}, S.~A. 1972, \solphys, 22, 344 \csname
  1972SoPh...22..344Slink\endcsname~\csname 1972SoPh...22..344Snote\endcsname

\bibitem[{{Spruit}(1977)}]{1977PhDT.......237S}
{Spruit}, H.~C. 1977, PhD thesis, Thesis University of Utrecht, The
  Netherlands. \csname 1977PhDT.......237Slink\endcsname~\csname
  1977PhDT.......237Snote\endcsname

\bibitem[{{Thomas}(1957)}]{1957ApJ...125..260T}
{Thomas}, R.~N. 1957, \apj, 125, 260 \csname
  1957ApJ...125..260Tlink\endcsname~\csname 1957ApJ...125..260Tnote\endcsname

\bibitem[{{Uitenbroek}(2000)}]{2000ApJ...536..481U}
{Uitenbroek}, H. 2000, \apj, 536, 481 \csname
  2000ApJ...536..481Ulink\endcsname~\csname 2000ApJ...536..481Unote\endcsname

\bibitem[{{Uitenbroek}(2001)}]{2001ApJ...557..389U}
{Uitenbroek}, H. 2001, \apj, 557, 389 \csname
  2001ApJ...557..389Ulink\endcsname~\csname 2001ApJ...557..389Unote\endcsname

\bibitem[{{Uitenbroek} \& {Bruls}(1992)}]{1992A&A...265..268U}
{Uitenbroek}, H. \& {Bruls}, J.~H.~M.~J. 1992, \aap, 265, 268 \csname
  1992A&A...265..268Ulink\endcsname~\csname 1992A&A...265..268Unote\endcsname

\bibitem[{{Vernazza} {et~al.}(1981){Vernazza}, {Avrett}, \&
  {Loeser}}]{1981ApJS...45..635V}
{Vernazza}, J.~E., {Avrett}, E.~H., \& {Loeser}, R. 1981, \apjs, 45, 635
  \csname 1981ApJS...45..635Vlink\endcsname~\csname
  1981ApJS...45..635Vnote\endcsname

\bibitem[{{Vitas} {et~al.}(2009){Vitas}, {Viticchi{\`e}}, {Rutten}, \&
  {V{\"o}gler}}]{2009A&A...499..301V}
{Vitas}, N., {Viticchi{\`e}}, B., {Rutten}, R.~J., \& {V{\"o}gler}, A. 2009,
  \aap, 499, 301 \csname 2009A&A...499..301Vlink\endcsname~\csname
  2009A&A...499..301Vnote\endcsname

\bibitem[{{Wedemeyer}(2016)}]{2016Msngr.163...15W}
{Wedemeyer}, S. 2016, The Messenger, 163, 15 \csname
  2016Msngr.163...15Wlink\endcsname~\csname 2016Msngr.163...15Wnote\endcsname

\end{thebibliography}

\end{small}

\end{document}